\documentclass[aps,prd,twocolumn,showpacs,superscriptaddress,nofootinbib,preprintnumbers]{revtex4-2}

\usepackage[pdftex,colorlinks]{hyperref}
\usepackage[dvipsnames]{xcolor}
\definecolor{phthaloblue}{rgb}{0.0, 0.06, 0.54}

\hypersetup{
    colorlinks=true,
    linkcolor=MidnightBlue,
    citecolor=MidnightBlue,
    filecolor=MidnightBlue,
    urlcolor=MidnightBlue,
    }
\usepackage[pdftex]{graphicx}
\usepackage{bm}
\usepackage{cancel}
\usepackage{here}
\usepackage{comment,braket}
\usepackage{epsf}
\usepackage{amsmath}
\usepackage{graphics}
\usepackage{amsfonts}
\usepackage{amssymb}
\usepackage{latexsym}
\usepackage{color}
\usepackage{natbib}
\usepackage[normalem]{ulem}
\usepackage{orcidlink}
\usepackage{cleveref}
\usepackage{physics}
\crefname{section}{Sec.}{Sec.}
\crefformat{section}{Sec.~#2#1#3}
\crefmultiformat{section}{Sec.~#2#1#3}{-#2#1#3}{, #2#1#3}{-#2#1#3}
\allowdisplaybreaks

\newcommand{\lettersection}[1]{\noindent{\bf \emph{#1.}}\,---\,}

\begin{document}

\preprint{FERMILAB-PUB-25-0016-T}

\title{Widen the Resonance: Probing a New Regime of Neutrino Self-Interactions \\
with Astrophysical Neutrinos}

\author{Isaac R. Wang\,\orcidlink{0000-0003-0789-218X}}
\email[Corresponding author:~]{isaacw@fnal.gov}
\affiliation{Theory Division, Fermi National Accelerator Laboratory, Batavia, Illinois 60510, USA}

\author{Xun-Jie Xu\,\orcidlink{0000-0003-3181-1386}}
\email[Corresponding author:~]{xuxj@ihep.ac.cn}
\affiliation{Institute of High Energy Physics, Chinese Academy of Sciences, Beijing 100049, China}

\author{Bei Zhou\,\orcidlink{0000-0003-1600-8835}}
\email[Corresponding author:~]{beizhou@fnal.gov}
\affiliation{Theory Division, Fermi National Accelerator Laboratory, Batavia, Illinois 60510, USA}
\affiliation{Kavli Institute for Cosmological Physics, University of Chicago, Chicago, Illinois 60637, USA}

\date{\today}

\begin{abstract}
Neutrino self-interactions beyond the standard model have profound implications in astrophysics and cosmology. In this Letter, we study an uncharted scenario in which one of the three neutrino species has a mass smaller than the temperature of the cosmic neutrino background. This results in a relativistic component that significantly broadens the absorption feature on the astrophysical neutrino spectra, in contrast to the sharply peaked absorption expected in the extensively studied scenarios assuming a fully nonrelativistic cosmic neutrino background. By solving the Boltzmann equations for neutrino absorption and regeneration, we demonstrate that this mechanism provides novel sensitivity to sub-keV mediator masses, well below the traditional $\sim 1$--100 MeV range. 
Future observations of the diffuse supernova neutrino background with Hyper-Kamiokande could probe coupling strengths down to $g \sim 10^{-8}$, surpassing existing constraints by orders of magnitude. These findings open new directions for discoveries and offer crucial insights into the interplay between neutrinos and the dark sector.
\end{abstract}

\bigskip
\maketitle

\lettersection{Introduction}Neutrinos, interacting feebly with matter, could strongly interact among themselves.
This intriguing possibility, which
is not only allowed by laboratory constraints~\cite{Bialynicka-Birula:1964ddi,Berryman:2018ogk,Blinov:2019gcj,Brdar:2020nbj,Deppisch:2020sqh,Dev:2024twk}
but also well motivated from various beyond-the-standard-model
theories~\cite{Chikashige:1980ui,Gelmini:1980re,Ma:2013yga,Lindner:2013awa,Berbig:2020wve,Xu:2020qek,Chauhan:2020mgv, Foroughi-Abari:2025upe},
has received growing interest in recent years---see, e.g., Ref.~\cite{Berryman:2022hds}
for a review. Moreover, the cosmological and astrophysical implications
of neutrino self-interactions are rich~\cite{Kreisch:2019yzn,RoyChoudhury:2020dmd,RoyChoudhury:2022rva,Venzor:2023aka,Das:2017iuj,Shalgar:2019rqe,Chang:2022aas,Fiorillo:2023ytr,Fiorillo:2023cas,Ng:2014pca,Ioka:2014kca,Bustamante:2020mep,Esteban:2021tub,Creque-Sarbinowski:2020qhz,Das:2022xsz,Akita:2022etk,Balantekin:2023jlg,Doring:2023vmk,Huang:2017egl,Luo:2020sho,Grohs:2020xxd,Li:2023puz,Wu:2023twu,Akita:2023iwq, Wang:2023csv, Kaplan:2024ydw},
with the most prominent ones related to cosmic microwave background (CMB) data interpretations~\cite{Kreisch:2019yzn,RoyChoudhury:2020dmd,RoyChoudhury:2022rva,Venzor:2023aka},
supernova dynamics~\cite{Das:2017iuj,Shalgar:2019rqe,Chang:2022aas,Fiorillo:2023ytr,Fiorillo:2023cas},
and astrophysical neutrino propagation~\cite{Ng:2014pca,Ioka:2014kca,Bustamante:2020mep,Esteban:2021tub,Creque-Sarbinowski:2020qhz,Das:2022xsz, Akita:2022etk,Balantekin:2023jlg, Doring:2023vmk}.

Astrophysical neutrinos propagating through cosmic distances could be attenuated due to scattering with the cosmic neutrino background (CNB) via neutrino self-interactions~\cite{Ng:2014pca,Ioka:2014kca,Bustamante:2020mep,Esteban:2021tub,Creque-Sarbinowski:2020qhz,Das:2022xsz,Akita:2022etk,Balantekin:2023jlg, Doring:2023vmk}.
To achieve observable attenuation, the mediator of neutrino self-interactions is often
assumed to be light, typically around $m_{\phi}\sim$ 1--100 MeV, 
with the coupling $g_{\nu}\sim 10^{-3}$--$10^{-1}$. 
The corresponding four-Fermi effective interaction strength is $G_{X}\equiv g^{2}/m_{\phi}^{2}\sim10^{5}G_{F}$ with $G_{F}$ being the Fermi constant. 
Below the MeV scale, cosmological observables from big bang nucleosynthesis (BBN) and the effective number of neutrino species ($N_{{\rm eff}}$) typically set much stronger constraints than astrophysical neutrinos and other probes.

In this \textit{Letter}, we propose an exceptionally interesting effect that could be exploited to probe a new regime of neutrino self-interactions \emph{well below the MeV scale}.
Compared to previous studies, our work takes into account a crucial factor that was neglected in the past but could substantially enhance the sensitivity reach, much stronger than the cosmological constraints. 
That is, the lightest neutrino species in the CNB today can still be relativistic, i.e., its mass being lower than the CNB temperature, which is allowed by all experimental data.

\begin{figure}[t]
    \centering    
    \includegraphics[width=0.49\textwidth]{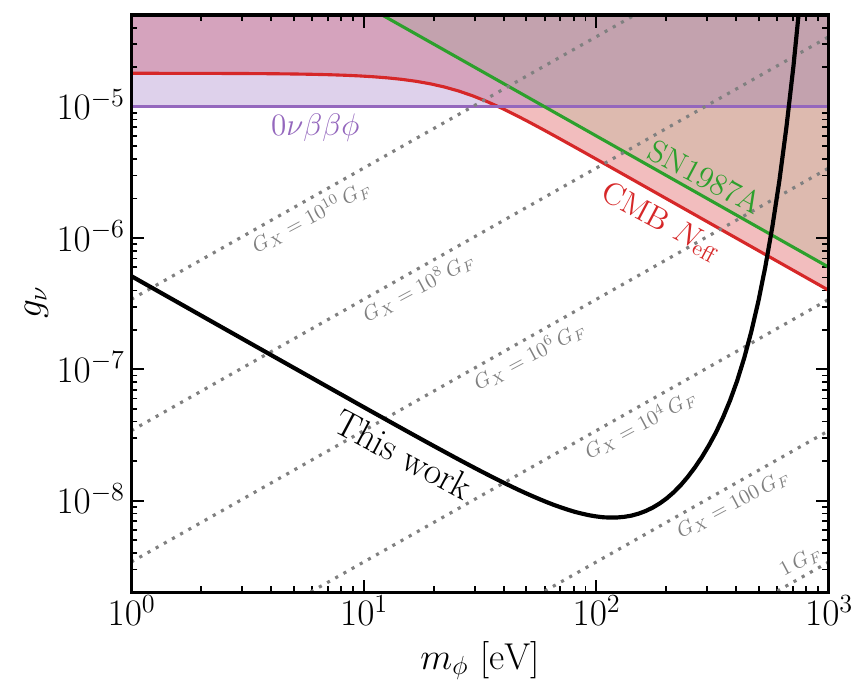}
    \caption{
        The parameter space of neutrino self-interactions in the low-mass regime. The observation of DSNB in upcoming neutrino detectors such as Hyper-Kamiokande can substantially improve current known bounds (colored regions)~\cite{Li:2023puz,Fiorillo:2022cdq,KamLAND-Zen:2012uen} 
        by a few orders of magnitude (the black line). 
        \protect\label{fig:final}
        }
\end{figure}

Fig.~\ref{fig:final} presents a preview of our results. 
We demonstrate that future observations of the diffuse supernova neutrino background 
(DSNB)~\cite{Horiuchi:2008jz,Beacom:2010kk} can probe $m_{\phi}$ of eV to sub-keV and $g_{\nu}$ down to $10^{-8}$.
This surpasses existing bounds by a few orders of magnitude. In terms of $G_{X}$,
 the best scenario may even reach ${\cal O}(100)G_{F}$.

Relativistic neutrinos may significantly enhance the absorption of astrophysical neutrinos during propagation.
This occurs at the resonance of the $s$-channel scattering, $\nu\nu\to\phi\to\nu\nu$,
where the mediator $\phi$ can be either on or off shell. In previous studies~\cite{Ng:2014pca,Ioka:2014kca,Bustamante:2020mep,Esteban:2021tub,Creque-Sarbinowski:2020qhz,Das:2022xsz,Akita:2022etk,Balantekin:2023jlg, Doring:2023vmk},
the CNB being scattered is assumed to be fully nonrelativistic.
Thus, the resonance occurs only in a very narrow energy range of astrophysical neutrinos.
In contrast, for relativistic CNB neutrinos that follow a thermal momentum distribution, the resonance can be reached
at much wider energies. 
This is because each astrophysical neutrino with a certain momentum can always find a CNB neutrino with appropriate momentum such that their Mandelstam variable $s$ matches $m_{\phi}^{2}$. 
In other words, \emph{relativistic CNB neutrinos widen the resonance}, which is the key observation made in this Letter.

The relativistic scenario becomes
even more motivated after the recent DESI publication reporting the
latest cosmological bound on the sum of neutrino masses~\cite{DESI:2024mwx, Herold:2024nvk},
as the upper bound is approaching the minimal value compatible with
neutrino oscillation data~\cite{Esteban:2024eli}\footnote{
If the cosmological bound further descends and becomes in tension
with the minimal value  (see e.g.~\cite{Wang:2024hen,Craig:2024tky}), then it
might suggest that some mechanisms~\cite{Fardon:2003eh,Kaplan:2004dq,Barger:2005mn,Cirelli:2005sg}
render cosmic neutrinos lighter than those involved in oscillation
measurements. In this case, more neutrino species in the CNB could
be relativistic.}.

In the subsequent analysis, we estimate the absorption rate of DSNB neutrinos
by the relativistic CNB to provide a qualitative discussion, highlighting the advantage of using such a scenario to probe neutrino self-interactions.
To quantitatively study the impact on the DSNB, we perform a numerical computation
by solving, for the first time, the Boltzmann equation for the coupled $\nu$-$\phi$ system instead of that for $\nu$ only. 

\begin{figure*}[t]
    \centering
\includegraphics[width=0.99\textwidth]{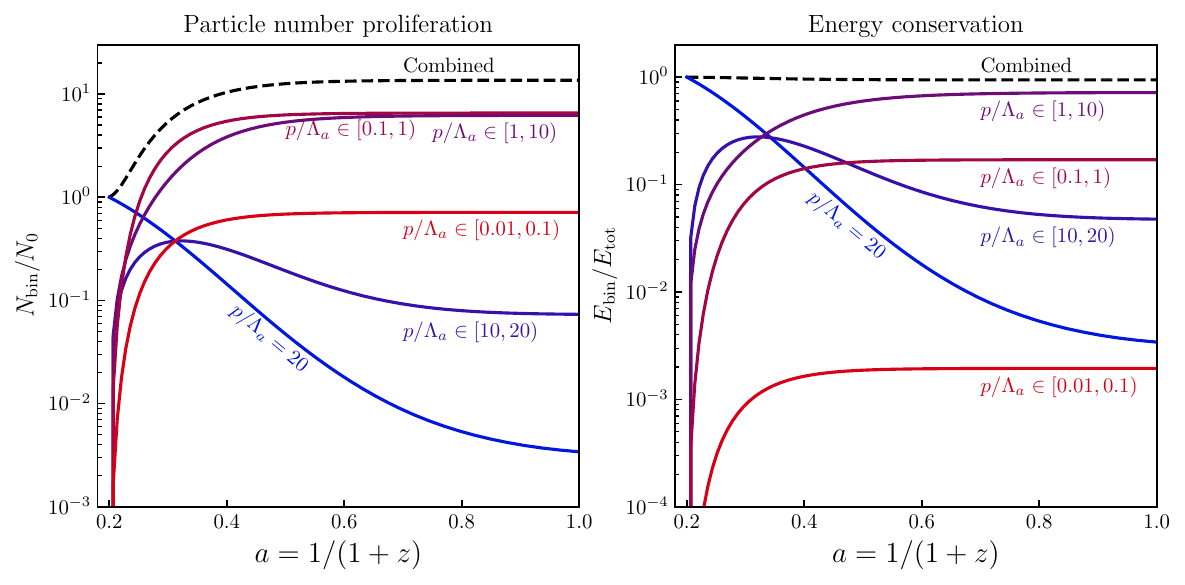}
\caption{Evolution of particle numbers (left) and energy (right), assuming $m_{\phi}=100$ eV, $g_{\nu}=6\times10^{-9}$,
and a monochromatic and instantaneous source emitting neutrinos with
$E_{\nu}=20$ MeV  at $z=4$. Here, the monotonically decreasing curves represent neutrinos that retain their initial comoving momentum, corresponding to $p/\Lambda_a=20$ 
where $\Lambda_a \equiv {\rm MeV}\cdot a_{i}/a$ and $a_i = 0.2$. Other colored curves represent particles that are subsequently produced. The black dashed lines represent the total number and energy of these particles.
}
\label{fig:conservation}
\end{figure*}

\lettersection{Interaction and mean-free path}We consider the following
Lagrangian for neutrino self-interactions: 
\begin{align}
\mathcal{L}\supset\frac{1}{2}(\partial\phi)^{2}-\frac{1}{2}m_{\phi}^{2}\phi^{2}+g_{\nu}\phi(\nu\nu+\nu^{\dagger}\nu^{\dagger})\,,\label{eq:L}
\end{align}
where $\phi$ is a real scalar and $\nu$ represents the two-component
Weyl-van der Waerden spinor of a neutrino. Hence, Majorana neutrinos
are implied by the Lagrangian.
Like the Majorana mass terms of neutrinos, the above interaction violates the lepton number, implying that antineutrinos can be produced from neutrinos via, e.g., $\nu\,\nu\to\phi \to\overline{\nu}\,\overline{\nu}$. Since both the source (DSNB) and the target (CNB) are approximately symmetric under $\nu\leftrightarrow\overline{\nu}$, our analysis below is also $\nu$-$\overline{\nu}$ symmetric, leading to $\nu$-$\overline{\nu}$ symmetric results,  unaffected by this lepton-number-violating feature.
For simplicity, we do not consider the flavor structure in this Letter. A full flavor analysis including the $\nu$-$\overline{\nu}$ difference  will be conducted in our upcoming work.

With the interaction in Eq.~\eqref{eq:L}, an astrophysical neutrino\footnote{Throughout this Letter, we use the term ``astrophysical neutrinos''
for neutrinos that acquire energies directly or indirectly from astrophysical sources.
This includes neutrinos both emitted directly from such sources and regenerated through self-interactions.
}, $\nu_{\rm A}$, propagating through the CNB may scatter off a neutrino in the CNB,
$\nu_{\rm C}$, via 
\begin{equation}
\nu_{\rm A}+\nu_{\rm C}\to X\thinspace,\ X\in\{2\nu_{\rm A},\ \phi,\ 2\phi,\ \cdots\}\thinspace.\label{eq:reaction}
\end{equation}
  It is useful to introduce the mean-free path of the propagation,
$L_{{\rm mean}}$. In the ultrarelativistic (UR) and nonrelativistic
(NR) scattering regimes, $L_{{\rm mean}}$ can be approximated as
\begin{equation}
L_{{\rm mean}} \simeq \begin{cases}
\left( \frac{g_{\nu}^{2}m_{\phi}^{2}T}{16\pi E_{\nu}^{2}}\exp\left[-\frac{m_{\phi}^{2}}{4TE_{\nu}}\right]+\cdots  \right)^{-1} & (\text{UR})\\
\left(n_{\rm C}\sigma_{{\rm NR}}\right)^{-1}  & (\text{NR})
\end{cases}\thinspace,\label{eq:mean}
\end{equation}
where $E_{\nu}$ denotes the energy of the incoming $\nu_{\rm A}$,
$T$ is the temperature of the CNB, $n_{\rm C}$ is the number density
of $\nu_{\rm C}$, and $\sigma_{{\rm NR}}$ denotes the NR cross section,
which is almost independent of the momentum of $\nu_{\rm C}$. The UR
and NR regimes correspond to the neutrino mass $m_{\nu}\ll T$ or
$m_{\nu} \gg T$, respectively. For more general cases, $L_{{\rm mean}}$
can be derived from collision terms of Boltzmann equations---see
the Supplemental Material. In the UR regime of Eq.~\eqref{eq:mean},
we only show the dominant contribution of $\nu_{\rm A}+\nu_{\rm C}\to\phi$,
with ``$\cdots$'' representing the remaining contributions.
While in the NR regime, the contribution of $\nu_{\rm A}+\nu_{\rm C}\to\phi$ to $\sigma_{{\rm NR}}$
reads:
\begin{equation}
\sigma_{{\rm NR}}^{({\rm res})}\simeq \pi g_\nu^2\delta(s-m_{\phi}^{2})\thinspace,\label{eq:delta}
\end{equation}
where $s \simeq 2 E_{\nu}m_{\nu}$.
Because of the $\delta$ function, Eq.~\eqref{eq:delta} gives rise to a very sharp
and narrow absorption of $\nu_{\rm A}$ by the CNB---see, e.g., Fig.~1
of Ref.~\cite{Ng:2014pca}. In practice, Eq.~\eqref{eq:delta} is
usually integrated into the resonance of $s$-channel scattering 
via the Breit-Wigner formalism (c.f.~Eq.~(25) in \cite{Tait:2008zz}),
rather than being calculated separately.

In contrast to the narrow absorption in the NR regime, the absorption in the UR regime is much wider.
As can be seen from Eq.~\eqref{eq:mean},  $L_{{\rm mean}}^{-1}$ in the UR regime peaks at $E_{\nu}\sim m_{\phi}^{2}/4T$, implying that $\nu_{\rm A}$ of this energy can be absorbed most efficiently. 
If $E_{\nu}$ varies around this value by a factor of a few, the absorption is still effective.

Therefore, by comparing the NR and UR regimes, we see that \emph{the most
crucial difference is a very sharp and narrow absorption versus
a much wider one.}

Taking the UR result with the present CNB temperature $T\simeq1.9$
K and a few benchmark values indicated below, we find 
\begin{equation}
\frac{L_{\rm mean}}{1\ \text{Gpc}} \simeq 0.8 \left( \frac{10^{-8}}{g_\nu} \cdot \frac{E_\nu}{20~\rm MeV} \cdot  \frac{0.1~\rm keV}{m_\phi}\right)^2 e^{\lambda}\thinspace,\label{eq:L-value}
\end{equation}
where $\lambda\equiv\frac{m_{\phi}^{2}}{4TE_{\nu}}$ and  for the
above benchmark values it gives $e^{\lambda}\sim{\cal O}(1)$. 
Obviously, the obtained $L_{{\rm mean}}$  for this benchmark is
much shorter than the radius of the observable universe, $R_{{\rm universe}}\simeq14$
Gpc. This implies that the relativistic CNB over the entire universe
would be opaque to a 20 MeV neutrino in the benchmark scenario! 
Therefore, observations of astrophysical neutrinos at such energies
from sources that are cosmologically distant should be able to probe
a sub-keV neutrino self-interaction mediator with $g_{\nu}\sim10^{-8}$.

The above estimate of the mean free path serves as a qualitative study
and has neglected the cosmological redshift. If $\nu_{\rm A}$ is produced
from high-redshift sources (e.g., redshift~$z\sim4-5$), it would propagate
through a much denser and more energetic (hence more likely to be
relativistic) neutrino background than the present CNB. A quantitative study, taking the cosmological redshift and other effects such as
the production of $\phi$ and secondary $\nu_{\rm A}$ into account, requires
solving the Boltzmann equation, as we discuss below.

\lettersection{Solving the Boltzmann equation}The Boltzmann equation
that governs the evolution of the phase space distributions of $\nu$
and $\phi$ in the expanding universe reads:
\begin{equation}
\left(\partial_{t}-Hp\partial_{p}\right)\left[\begin{array}{c}
f_{\nu}\\[2mm]
f_{\phi}
\end{array}\right]=\left[\begin{array}{c}
-f_{\nu}\Gamma_{\nu}^{-}+(1-f_{\nu})\left(\Gamma_{\nu}^{+}+{\cal S}_{\nu}\right)\\[2mm]
-f_{\phi}\Gamma_{\phi}^{-}+(1+f_{\phi})\Gamma_{\phi}^{+}
\end{array}\right],\label{eq:Boltzmann}
\end{equation}
where $f_{x}$ is the phase space distribution of species $x=\nu$
or $\phi$, $H=a^{-1}da/dt$ is the Hubble parameter with $a$ the
scale factor, $\Gamma_{x}^{+}$ and $\Gamma_{x}^{-}$ are the production
and depletion rates of $x$ via particle physics processes such as
those in Eq.~\eqref{eq:reaction}, and ${\cal S}_{\nu}$ accounts
for the production of $\nu$ from astrophysical sources.

In principle, $f_{\nu}$ could include both astrophysical 
and CNB neutrinos, with the former being viewed as a high-energy nonthermal addition to the thermal spectrum of the latter. However, the very
hierarchical energy regimes between them and the low number density
of the former make this treatment impractical. Hence, we separate
the high-energy part from the thermal part and consider $f_{\nu}$
as the phase space distribution of $\nu_{\rm A}$ only. Under this convention,
we have $f_{\nu,\phi}\ll1$, which implies that the Pauli-blocking
and Bose-enhancement factors $1\mp f_{\nu,\phi}$ can be neglected. 

In the UR regime, the dominant processes contributing to $\Gamma_{\nu,\phi}^{\pm}$
are $\nu_{\rm A}+\nu_{\rm C}\to\phi$ and $\phi\to\nu_{\rm A}+\nu_{\rm A}$, with
their contributions proportional to $g_{\nu}^{2}$. Other processes
such as $\nu_{\rm A}+\nu_{\rm C}\to\nu_{\rm A}+\nu_{\rm A}$ (without resonance) and
$\nu_{\rm A}+\nu_{\rm C}\to\phi+\phi$ are subdominant in the regime of our
interest because their contributions are suppressed by $g_{\nu}^{4}$.
In this Letter, we focus on the ${\cal O}(g_{\nu}^{2})$ contributions,
which can be calculated analytically (see Sec.S1 in the Supplemental Material for the derivations):
\begin{align}
\Gamma_{\nu}^{-} & =\frac{g_{\nu}^{2}m_{\phi}^{2}T}{16\pi E_{\nu}^{2}}\exp\left[-\frac{m_{\phi}^{2}}{4TE_{\nu}}\right],\label{eq:-10}\\
\Gamma_{\nu}^{+} & =\frac{2g_{\nu}^{2}m_{\phi}^{2}}{16\pi E_{\nu}^{2}}\int_{E_{\phi}^{\min}}^{\infty}{\rm d}E_{\phi}f_{\phi}\thinspace,\label{eq:-11}\\
\Gamma_{\phi}^{+} & =\frac{g_{\nu}^{2}m_{\phi}^{2}}{16\pi E_{\phi}p_{\phi}}\int_{E_{\nu}^{-}}^{E_{\nu}^{+}}dE_{\nu}f_{\nu}\exp\left[-\frac{E_{\phi}-p_{\nu}}{T}\right],\label{eq:-12}\\
\Gamma_{\phi}^{-} & =\frac{g_{\nu}^{2}m_{\phi}^{2}}{16\pi E_{\phi}}\thinspace,\label{eq:-13}
\end{align}
where $E_{\phi}$ and $p_{\phi}$ denote the energy and momentum of
$\phi$, $E_{\phi}^{\min}=\frac{m_{\phi}^{2}}{4E_{\nu}}+E_{\nu}$,
and $E_{\nu}^{\mp}=(E_{\phi}\mp p_{\phi})/2$. 
With the analytical expressions of $\Gamma_{\nu,\phi}^{\pm}$ and
a given source ${\cal S}_{\nu}$, it is straightforward to solve the
Boltzmann equation numerically.

Fig.~\ref{fig:conservation} shows an example of a monochromatic and instantaneous source emitting neutrinos with
$E_{\nu}=20$ MeV at $z=4$.
In addition, we set $m_{\phi}=100$ eV, $g_{\nu}=6\times10^{-9}$ for neutrino propagation.
The left and right panels show how the particle
number and energy in each given momentum bin evolve, respectively. Here the particle
number and energy in a bin are defined as $N_{{\rm bin}}\equiv\int_{{\rm bin}}\tilde{f}_{\nu+\phi}(\tilde{p})\frac{d^{3}\tilde{p}}{(2\pi)^{3}}$
and $E_{{\rm bin}}\equiv\int_{{\rm bin}}\tilde{f}_{\nu+\phi}(\tilde{p})\tilde{p}\frac{d^{3}\tilde{p}}{(2\pi)^{3}}$,
where $\tilde{p}=ap$ is the comoving momentum and $\tilde{f}_{\nu+\phi}(\tilde{p})\equiv f_{\nu}(p)+f_\phi(p)$.
Note that our numerical computation throughout the Letter uses much finer momentum binning than what is shown in the figure.

As is shown in the left panel, the total number of astrophysical neutrinos
(black dashed line) increases rapidly after the initial production
from the source.
This is because $\nu_{\rm A}+\nu_{\rm C}\to\phi$ and $\phi\to\nu_{\rm A}+\nu_{\rm A}$
proceed efficiently at this stage. The two processes together lead
to an exponential growth of $\nu_{\rm A}$ with degraded energies, and
meanwhile deplete $\nu_{\rm A}$ at the initial energy. 
Despite the proliferation
of the particle number, the total energy is approximately conserved
(up to proper comoving factors), as is shown by the black dashed line in the right panel. The energy conservation relies on two approximations:
(i) the energy contribution from $\nu_{\rm C}$ is negligible, and (ii) the involved species are all relativistic\footnote{Otherwise, there would be the dilution-resistant effect caused by
the mass~\cite{Li:2023puz}.}. The validity of (i) is guaranteed by the low temperature of CNB;
and the validity of (ii) is implied by $E_{\nu}\gg m_{\phi}$. From the energy conservation and the particle number proliferation, we anticipate that the most prominent effect of neutrino self-interactions on the DSNB should be a substantial enhancement of its low-energy
spectrum in combination with significant absorptions at high energies.

\lettersection{Impact on the DSNB spectrum}To reach the resonance
in the UR regime, $E_{\nu}$ needs to be around 1--100 MeV, which
roughly matches the energy range of supernova neutrinos. Hence, the most suitable neutrino source to be considered in this work would be supernovae, provided that their distances are at the cosmological scale (Gpc).   Although the neutrino flux of a single supernova
at such a large distance is far from being observable in current and future experiments, the combined neutrino flux of all supernovae in the entire universe (i.e., DSNB) is likely to be detected in the
foreseeable future~\cite{Beacom:2010kk}. Therefore, it is tempting to investigate whether
and how the DSNB might be altered in the new physics scenario considered
here. 

\begin{figure}
\centering
\includegraphics[width=0.49\textwidth]{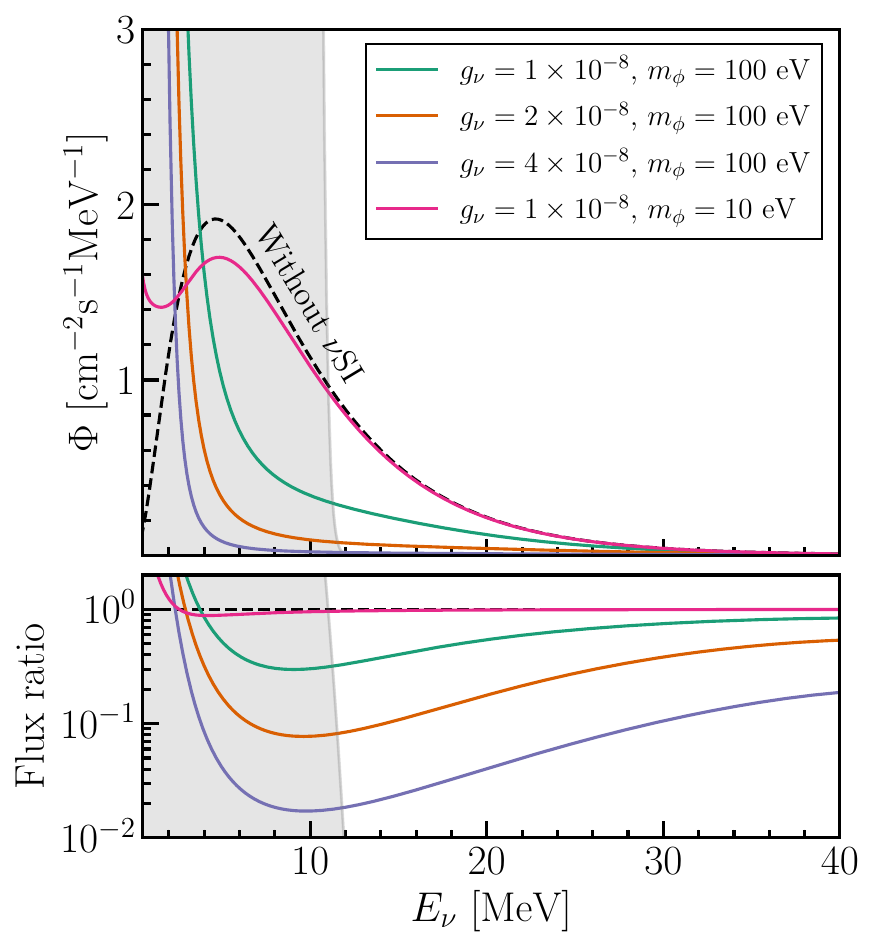}
\caption{The impact of neutrino self-interactions ($\nu$SI) on the DSNB energy spectrum. \textit{Upper panel}: neutrino flux. \textit{Lower panel}: neutrino flux normalized by the result from the propagation of DSNB neutrinos without $\nu$SI.
The gray-shaded region indicates the background from reactor neutrinos~\cite{Baldoncini:2014vda,Vitagliano:2019yzm}.
The black dashed curves indicate the results without $\nu$SI.
The DSNB temperature is assumed to be 6 MeV here.
}
\label{fig:DSNB}
\end{figure}

 The source term of the DSNB can be computed by~\cite{Horiuchi:2008jz, Beacom:2010kk},
\begin{align}
\mathcal{S}_{\nu}(E_{\nu},a)=\frac{2\pi^{2}}{E_{\nu}^{2}}\cdot\frac{dN}{dE_{\nu}}\cdot\frac{R_{{\rm SN}}}{a^{3}}\thinspace,\label{eq:luminosity}
\end{align}
where $dN/dE_{\nu}$ denotes the neutrino emission per supernova and
$R_{{\rm SN}}$ is the rate of core-collapse supernova. Both can be
approximated by analytical expressions and have been provided by Ref.~\cite{Horiuchi:2008jz}.
Following the same setup in Ref.~\cite{Horiuchi:2008jz}, we calculate
Eq.~\eqref{eq:luminosity} and substitute it into the Boltzmann equation.
Solving the equation, we obtain the phase space distribution $f_{\nu}$,
which can be recast into the neutrino flux $\Phi_{\nu}$ by
\begin{equation}
\Phi_{\nu}=a^{3}\frac{E_{\nu}^{2}}{2\pi^{2}}f_{\nu}\thinspace.\label{eq:flux}
\end{equation} 

Fig.~\ref{fig:DSNB} shows the impact on the DSNB spectrum from neutrino self-interaction.
In the upper panel, the black dashed curve represents the DSNB neutrino flux without neutrino self-interactions, assuming a temperature of 6 MeV.
This curve is confirmed to align with Fig.~4 in Refs.~\cite{Horiuchi:2008jz,Beacom:2010kk}
and is also approximately consistent with Fig.~1 in \cite{Vitagliano:2019yzm}.
The colored curves depict the results of neutrino self-interaction.
We show the result with three different couplings, $g=10^{-8}$, $g=2 \times 10^{-8}$, and $g = 4 \times 10^{-8}$ for $m_\phi = 100~\rm eV$, while also showing the result of $g=10^{-8}$ for $m_\phi = 10~\rm eV$ to demonstrate the impact of the mediator mass.
The lower panel illustrates the normalized results compared to free propagation, highlighting both creation and absorption effects.
In all cases with the interaction, the neutrino flux in the MeV range is absorbed by $\nu_{\rm A} + \nu_{\rm C} \to \phi$.
One can see that the absorption in the lower panel behaves as a \textit{wide dip}, indicating the result of the \textit{widened resonance}.
Next, the produced $\phi$ particles decay into two neutrinos, with each roughly carrying half of the energy of $\nu_{\rm A}$, given that CNB temperature is lower than $\nu_{\rm A}$ energy by a few orders of magnitude.
These created neutrinos further scatter with CNB to create more $\phi$ particles, which then decay into pairs of neutrinos with halved energies.
Consequently, there is an accumulation of neutrinos in the low-energy spectrum, leading to a blowup behavior in flux in the lower-energy range, as shown in Fig.~\ref{fig:DSNB}.

\lettersection{Detection and $\chi^{2}$ analysis}For DSNB detection, we focus on the dominant inverse beta decay (IBD) channel\footnote{Other neutrino detectors such as Deep Underground Neutrino Experiment (DUNE)  and Jiangmen Underground Neutrino Observatory (JUNO) may have their own
advantages such as new detection channels other than IBD and the possibility of 
detecting the DSNB at lower energies.} in the upcoming Hyper-Kamiokande (HK) detector~\cite{Hyper-Kamiokande:2018ofw,Abe:2011ts} that is expected
to detect the DSNB with robust statistics due to its unprecedentedly large volume and excellent background reduction if it is Gadolinium doped.
With a $3740\ {\rm kt}\cdot{\rm yr}$ exposure, HK is expected to
detect $\sim800$ DSNB events in the standard scenario above 12~MeV,
while in the new physics scenario, the events above a certain energy
could be substantially reduced. 
The calculation of DSNB event rates at HK has been well-established in the literature---see, e.g., Ref.~\cite{Balantekin:2023jlg}. We detail it together with a discussion on the detection backgrounds in Sec.~S4 of the Supplemental Material.

To quantify the observability of the new physics effect, we adopt
the following $\chi^{2}$ function~\cite{ParticleDataGroup:2024cfk}:
\begin{equation}
\chi^{2}=\left(\frac{y}{\sigma_{y}}\right)^{2}+\sum_{i}\left(\frac{(1+y)N_{i}-N_{{\rm st},i}}{\sigma_{i}}\right)^{2},\label{eq:chi}
\end{equation}
where $N_{i}$ and $N_{{\rm st},i}$ denote the numbers of events in the $i$-th energy bin in the new physics and the standard model scenarios, respectively;
$y$ is a normalization factor; and $\sigma_{i}$ and $\sigma_{y}$
denote the uncertainties of $N_{i}$ and $y$, respectively. With sufficient statistics ($N_{{\rm st},i}\gg 1$), one can take $\sigma_{i} = \sqrt{N_{{\rm st},i}+N_{{\rm bkg},i}}$ where $N_{{\rm bkg},i}$ is the number of background events.
For the normalization uncertainty, we take $\sigma_{y}=0.5$, which
can be justified from DSNB calculations over the past ten years---see the Supplemental Material for detailed discussions. This uncertainty
is partially caused by the star formation rate~\cite{Horiuchi:2008jz}
and could be improved by future astrophysical observations. In our
analysis, we adopt the frequentist treatment to marginalize $y$~\cite{ParticleDataGroup:2024cfk}.
In addition, the DSNB temperature $T_{{\rm SN}}$ is uncertain and
may affect the spectral shape. We vary $T_{{\rm SN}}$ in $[4,\ 8]$
MeV and marginalize it as well.
Fig.~\ref{fig:final} presents our projected sensitivity reach. Our results indicate that observations of DSNB at HK can probe neutrino self-interactions with $m_{\phi}$ from eV to sub-keV and $g_{\nu}$ down to $10^{-8}$.

In the shown parameter space, there are a few existing bounds derived from 
cosmology~\cite{Sandner:2023ptm,Li:2023puz}, 
the detection of SN1987A neutrinos~\cite{Choi:1989hi,Kachelriess:2000qc,Farzan:2002wx,Fiorillo:2022cdq,Vogl:2024ack}, 
and neutrinoless double beta decay experiments searching for majoron emissions ($0\nu\beta\beta\phi$)~\cite{KamLAND-Zen:2012uen,Brune:2018sab,Blum:2018ljv,Deppisch:2020sqh,GERDA:2022ffe}.
The cosmological bound relies on the production of $\phi$ in the early universe via $\nu\nu\to \phi$ and $\nu\nu\to \phi\phi$, 
with the production rates proportional to $g_{\nu}^2$ and $g_{\nu}^4$, respectively. In the low-mass regime, the former is suppressed by the mass of $\phi$, causing the latter to dominate the production of very light $m_{\phi}$. 
After neutrino decoupling, the produced $\phi$ and their subsequent decay products may affect the Hubble expansion rate, thereby altering BBN and CMB predictions. This effect can be quantified by the cosmological parameter $N_{{\rm eff}}$. 
Here we take the $N_{{\rm eff}}$ bound on neutrinophilic scalars from Ref.~\cite{Li:2023puz}, assuming the allowed deviation
of $N_{{\rm eff}}$ is less than $0.285$~\cite{Planck:2018vyg}. 
For the SN1987A bound, we adopt the most stringent one from Ref.~\cite{Fiorillo:2022cdq}.
As for the $0\nu\beta\beta\phi$ bound, since $m_{\phi}$ here is
well below the $Q$ value of the process, it is almost independent
of $m_{\phi}$. Experimental searches for $0\nu\beta\beta\phi$ typically set an upper bound
on $g_{\nu}$ around $10^{-5}$~\cite{KamLAND-Zen:2012uen,GERDA:2022ffe}.  

Comparing our result with the existing bounds in Fig.~\ref{fig:final}, we see that a substantial part of the unexplored parameter space can be probed by DSNB observations. In this work, we focus on supernova neutrinos mainly because the energy scale at ${\cal O}(10)$ MeV is ideal for the resonant production of sub-keV $\phi$. Nevertheless, our calculation can be readily applied to other astrophysical sources at much higher energies. For instance, PeV or EeV neutrinos would allow us to explore the widened resonance for $m_{\phi}\sim 1$ or $30$ MeV, respectively. We leave these interesting scenarios to future work.

\lettersection{Summary and conclusions}Strong beyond-the-standard-model neutrino self-interactions are well motivated and exhibit rich phenomena in astrophysics and cosmology.
An intriguing consequence is that astrophysical neutrinos propagating through cosmic distances could be attenuated due to scattering with the 
CNB, typically resulting in a sharp absorption feature on the spectrum due to the resonance production of the mediator from the self-interaction.
This mechanism provides stringent constraints on light mediators with masses of $\sim 1$--100 MeV and coupling strengths in the range of $10^{-3}$ to $10^{-1}$.

In this letter, we consider an uncharted scenario in which the absorption is broadened significantly due to a relativistic component of the CNB. This occurs if the mass of the lightest neutrino species is below the CNB temperature,
which is allowed by neutrino-oscillation measurements and motivated by recent DESI data; thus, the CNB could follow a wide thermal momentum distribution, enabling a much wider range of astrophysical neutrino energies to satisfy the resonance condition. 

We demonstrate that this broadened absorption feature offers a novel sensitivity to sub-keV mediator masses, well below the traditional 1--100~MeV range. By solving the Boltzmann equations for the coupled neutrino-mediator system, we quantitatively evaluate the absorption and regeneration of DSNB neutrinos during propagation. Our results show that for mediator masses in the eV to sub-keV range, future observations of the DSNB by Hyper-Kamiokande would have a sensitivity of the coupling strength up to $g \sim 10^{-8}$, surpassing existing constraints by a few orders of magnitude.

This scenario warrants various studies in future work. For example, the proliferation of sub-10~MeV neutrinos could exceed the fluxes of neutrinos from reactors and other sources. Furthermore, rich phenomenology could manifest in TeV--EeV astrophysical neutrinos observed by IceCube and other observatories. Altogether, they will significantly extend our reach in the parameter space of neutrino self-interactions,
offering valuable insights into the mystery of neutrinos and their possible interplay with the dark sector.

{\bf Acknowledgement}
We are grateful to John Beacom and Pedro Machado for their helpful discussions.
I.\,R.\,W. and B.\,Z. were supported by Fermi Research Alliance, LLC under Contract No.~DE-AC02-07CH11359 with the U.S. Department of Energy, Office of Science, Office of High Energy Physics, and are currently supported by Fermi Forward Discovery Group, LLC under Contract No. 89243024CSC000002 with the U.S. Department of Energy, Office of Science, Office of High Energy Physics.
I.\,R.\,W. is also supported by DOE Distinguished Scientist Fellowship Grant No. FNAL 22-33. The work of X.\,J.\,X. is supported in part by the National Natural Science Foundation of China (NSFC) under Grant No.~12141501 and also by the CAS Project for Young Scientists in Basic Research (YSBR-099).

\bibliographystyle{utphys}
\bibliography{main}

\providecommand{\href}[2]{#2}\begingroup\raggedright\begin{thebibliography}{10}

\bibitem{Bialynicka-Birula:1964ddi}
Z.~Bialynicka-Birula, ``{Do Neutrinos Interact between Themselves?},''
  \href{http://dx.doi.org/10.1007/BF02749481}{{\em Nuovo Cim.} {\bf 33} (1964)
  1484--1487}.

\bibitem{Berryman:2018ogk}
J.~M. Berryman, A.~De~Gouv\^ea, K.~J. Kelly, and Y.~Zhang,
  ``{Lepton-Number-Charged Scalars and Neutrino Beamstrahlung},''
  \href{http://dx.doi.org/10.1103/PhysRevD.97.075030}{{\em Phys. Rev. D} {\bf
  97} (2018) no.~7, 075030}, \href{http://arxiv.org/abs/1802.00009}{{\tt
  arXiv:1802.00009 [hep-ph]}}.

\bibitem{Blinov:2019gcj}
N.~Blinov, K.~J. Kelly, G.~Z. Krnjaic, and S.~D. McDermott, ``{Constraining the
  Self-Interacting Neutrino Interpretation of the Hubble Tension},''
  \href{http://dx.doi.org/10.1103/PhysRevLett.123.191102}{{\em Phys. Rev.
  Lett.} {\bf 123} (2019) no.~19, 191102},
  \href{http://arxiv.org/abs/1905.02727}{{\tt arXiv:1905.02727 [astro-ph.CO]}}.

\bibitem{Brdar:2020nbj}
V.~Brdar, M.~Lindner, S.~Vogl, and X.-J. Xu, ``{Revisiting neutrino
  self-interaction constraints from $Z$ and $\tau$ decays},''
  \href{http://dx.doi.org/10.1103/PhysRevD.101.115001}{{\em Phys. Rev. D} {\bf
  101} (2020) no.~11, 115001}, \href{http://arxiv.org/abs/2003.05339}{{\tt
  arXiv:2003.05339 [hep-ph]}}.

\bibitem{Deppisch:2020sqh}
F.~F. Deppisch, L.~Graf, W.~Rodejohann, and X.-J. Xu, ``{Neutrino
  Self-Interactions and Double Beta Decay},''
  \href{http://dx.doi.org/10.1103/PhysRevD.102.051701}{{\em Phys. Rev. D} {\bf
  102} (2020) no.~5, 051701}, \href{http://arxiv.org/abs/2004.11919}{{\tt
  arXiv:2004.11919 [hep-ph]}}.

\bibitem{Dev:2024twk}
P.~S.~B. Dev, D.~Kim, D.~Sathyan, K.~Sinha, and Y.~Zhang, ``{New Laboratory
  Constraints on Neutrinophilic Mediators},''
  \href{http://arxiv.org/abs/2407.12738}{{\tt arXiv:2407.12738 [hep-ph]}}.

\bibitem{Chikashige:1980ui}
Y.~Chikashige, R.~N. Mohapatra, and R.~D. Peccei, ``{Are There Real Goldstone
  Bosons Associated with Broken Lepton Number?},''
  \href{http://dx.doi.org/10.1016/0370-2693(81)90011-3}{{\em Phys. Lett. B}
  {\bf 98} (1981)  265--268}.

\bibitem{Gelmini:1980re}
G.~B. Gelmini and M.~Roncadelli, ``{Left-Handed Neutrino Mass Scale and
  Spontaneously Broken Lepton Number},''
  \href{http://dx.doi.org/10.1016/0370-2693(81)90559-1}{{\em Phys. Lett. B}
  {\bf 99} (1981)  411--415}.

\bibitem{Ma:2013yga}
E.~Ma, I.~Picek, and B.~Radov\v{c}i\'c, ``{New Scotogenic Model of Neutrino
  Mass with $U(1)_D$ Gauge Interaction},''
  \href{http://dx.doi.org/10.1016/j.physletb.2013.09.049}{{\em Phys. Lett. B}
  {\bf 726} (2013)  744--746}, \href{http://arxiv.org/abs/1308.5313}{{\tt
  arXiv:1308.5313 [hep-ph]}}.

\bibitem{Lindner:2013awa}
M.~Lindner, D.~Schmidt, and A.~Watanabe, ``{Dark matter and U(1)' symmetry for
  the right-handed neutrinos},''
  \href{http://dx.doi.org/10.1103/PhysRevD.89.013007}{{\em Phys. Rev. D} {\bf
  89} (2014) no.~1, 013007}, \href{http://arxiv.org/abs/1310.6582}{{\tt
  arXiv:1310.6582 [hep-ph]}}.

\bibitem{Berbig:2020wve}
M.~Berbig, S.~Jana, and A.~Trautner, ``{The Hubble tension and a renormalizable
  model of gauged neutrino self-interactions},''
  \href{http://dx.doi.org/10.1103/PhysRevD.102.115008}{{\em Phys. Rev. D} {\bf
  102} (2020) no.~11, 115008}, \href{http://arxiv.org/abs/2004.13039}{{\tt
  arXiv:2004.13039 [hep-ph]}}.

\bibitem{Xu:2020qek}
X.-J. Xu, ``{The $\nu_{R}$-philic scalar: its loop-induced interactions and
  Yukawa forces in LIGO observations},''
  \href{http://dx.doi.org/10.1007/JHEP09(2020)105}{{\em JHEP} {\bf 09} (2020)
  105}, \href{http://arxiv.org/abs/2007.01893}{{\tt arXiv:2007.01893
  [hep-ph]}}.

\bibitem{Chauhan:2020mgv}
G.~Chauhan and X.-J. Xu, ``{How dark is the $\nu_R$-philic dark photon?},''
  \href{http://dx.doi.org/10.1007/JHEP04(2021)003}{{\em JHEP} {\bf 04} (2021)
  003}, \href{http://arxiv.org/abs/2012.09980}{{\tt arXiv:2012.09980
  [hep-ph]}}.

\bibitem{Foroughi-Abari:2025upe}
S.~Foroughi-Abari, K.~J. Kelly, M.~Rai, and Y.~Zhang, ``{Enabling Strong
  Neutrino Self-Interaction with an Unparticle Mediator},''
  \href{http://dx.doi.org/10.1103/PhysRevLett.134.181001}{{\em Phys. Rev.
  Lett.} {\bf 134} (2025) no.~18, 181001},
  \href{http://arxiv.org/abs/2501.02049}{{\tt arXiv:2501.02049 [hep-ph]}}.

\bibitem{Berryman:2022hds}
J.~M. Berryman {\em et al.}, ``{Neutrino self-interactions: A white paper},''
  \href{http://dx.doi.org/10.1016/j.dark.2023.101267}{{\em Phys. Dark Univ.}
  {\bf 42} (2023)  101267}, \href{http://arxiv.org/abs/2203.01955}{{\tt
  arXiv:2203.01955 [hep-ph]}}.

\bibitem{Kreisch:2019yzn}
C.~D. Kreisch, F.-Y. Cyr-Racine, and O.~Dor\'e, ``{Neutrino puzzle: Anomalies,
  interactions, and cosmological tensions},''
  \href{http://dx.doi.org/10.1103/PhysRevD.101.123505}{{\em Phys. Rev. D} {\bf
  101} (2020) no.~12, 123505}, \href{http://arxiv.org/abs/1902.00534}{{\tt
  arXiv:1902.00534 [astro-ph.CO]}}.

\bibitem{RoyChoudhury:2020dmd}
S.~Roy~Choudhury, S.~Hannestad, and T.~Tram, ``{Updated constraints on massive
  neutrino self-interactions from cosmology in light of the $H_0$ tension},''
  \href{http://dx.doi.org/10.1088/1475-7516/2021/03/084}{{\em JCAP} {\bf 03}
  (2021)  084}, \href{http://arxiv.org/abs/2012.07519}{{\tt arXiv:2012.07519
  [astro-ph.CO]}}.

\bibitem{RoyChoudhury:2022rva}
S.~Roy~Choudhury, S.~Hannestad, and T.~Tram, ``{Massive neutrino
  self-interactions and inflation},''
  \href{http://dx.doi.org/10.1088/1475-7516/2022/10/018}{{\em JCAP} {\bf 10}
  (2022)  018}, \href{http://arxiv.org/abs/2207.07142}{{\tt arXiv:2207.07142
  [astro-ph.CO]}}.

\bibitem{Venzor:2023aka}
J.~Venzor, G.~Garcia-Arroyo, J.~De-Santiago, and A.~P\'erez-Lorenzana,
  ``{Resonant neutrino self-interactions and the H0 tension},''
  \href{http://dx.doi.org/10.1103/PhysRevD.108.043536}{{\em Phys. Rev. D} {\bf
  108} (2023) no.~4, 043536}, \href{http://arxiv.org/abs/2303.12792}{{\tt
  arXiv:2303.12792 [astro-ph.CO]}}.

\bibitem{Das:2017iuj}
A.~Das, A.~Dighe, and M.~Sen, ``{New effects of non-standard self-interactions
  of neutrinos in a supernova},''
  \href{http://dx.doi.org/10.1088/1475-7516/2017/05/051}{{\em JCAP} {\bf 05}
  (2017)  051}, \href{http://arxiv.org/abs/1705.00468}{{\tt arXiv:1705.00468
  [hep-ph]}}.

\bibitem{Shalgar:2019rqe}
S.~Shalgar, I.~Tamborra, and M.~Bustamante, ``{Core-collapse supernovae stymie
  secret neutrino interactions},''
  \href{http://dx.doi.org/10.1103/PhysRevD.103.123008}{{\em Phys. Rev. D} {\bf
  103} (2021) no.~12, 123008}, \href{http://arxiv.org/abs/1912.09115}{{\tt
  arXiv:1912.09115 [astro-ph.HE]}}.

\bibitem{Chang:2022aas}
P.-W. Chang, I.~Esteban, J.~F. Beacom, T.~A. Thompson, and C.~M. Hirata,
  ``{Toward Powerful Probes of Neutrino Self-Interactions in Supernovae},''
  \href{http://dx.doi.org/10.1103/PhysRevLett.131.071002}{{\em Phys. Rev.
  Lett.} {\bf 131} (2023) no.~7, 071002},
  \href{http://arxiv.org/abs/2206.12426}{{\tt arXiv:2206.12426 [hep-ph]}}.

\bibitem{Fiorillo:2023ytr}
D.~F.~G. Fiorillo, G.~G. Raffelt, and E.~Vitagliano, ``{Large Neutrino Secret
  Interactions Have a Small Impact on Supernovae},''
  \href{http://dx.doi.org/10.1103/PhysRevLett.132.021002}{{\em Phys. Rev.
  Lett.} {\bf 132} (2024) no.~2, 021002},
  \href{http://arxiv.org/abs/2307.15115}{{\tt arXiv:2307.15115 [hep-ph]}}.

\bibitem{Fiorillo:2023cas}
D.~F.~G. Fiorillo, G.~G. Raffelt, and E.~Vitagliano, ``{Supernova emission of
  secretly interacting neutrino fluid: Theoretical foundations},''
  \href{http://dx.doi.org/10.1103/PhysRevD.109.023017}{{\em Phys. Rev. D} {\bf
  109} (2024) no.~2, 023017}, \href{http://arxiv.org/abs/2307.15122}{{\tt
  arXiv:2307.15122 [hep-ph]}}.

\bibitem{Ng:2014pca}
K.~C.~Y. Ng and J.~F. Beacom, ``{Cosmic neutrino cascades from secret neutrino
  interactions},'' \href{http://dx.doi.org/10.1103/PhysRevD.90.065035}{{\em
  Phys. Rev. D} {\bf 90} (2014) no.~6, 065035},
  \href{http://arxiv.org/abs/1404.2288}{{\tt arXiv:1404.2288 [astro-ph.HE]}}.
  [Erratum: Phys.Rev.D 90, 089904 (2014)].

\bibitem{Ioka:2014kca}
K.~Ioka and K.~Murase, ``{IceCube PeV\textendash{}EeV neutrinos and secret
  interactions of neutrinos},''
  \href{http://dx.doi.org/10.1093/ptep/ptu090}{{\em PTEP} {\bf 2014} (2014)
  no.~6, 061E01}, \href{http://arxiv.org/abs/1404.2279}{{\tt arXiv:1404.2279
  [astro-ph.HE]}}.

\bibitem{Bustamante:2020mep}
M.~Bustamante, C.~Rosenstr\o{}m, S.~Shalgar, and I.~Tamborra, ``{Bounds on
  secret neutrino interactions from high-energy astrophysical neutrinos},''
  \href{http://dx.doi.org/10.1103/PhysRevD.101.123024}{{\em Phys. Rev. D} {\bf
  101} (2020) no.~12, 123024}, \href{http://arxiv.org/abs/2001.04994}{{\tt
  arXiv:2001.04994 [astro-ph.HE]}}.

\bibitem{Esteban:2021tub}
I.~Esteban, S.~Pandey, V.~Brdar, and J.~F. Beacom, ``{Probing secret
  interactions of astrophysical neutrinos in the high-statistics era},''
  \href{http://dx.doi.org/10.1103/PhysRevD.104.123014}{{\em Phys. Rev. D} {\bf
  104} (2021) no.~12, 123014}, \href{http://arxiv.org/abs/2107.13568}{{\tt
  arXiv:2107.13568 [hep-ph]}}.

\bibitem{Creque-Sarbinowski:2020qhz}
C.~Creque-Sarbinowski, J.~Hyde, and M.~Kamionkowski, ``{Resonant neutrino
  self-interactions},''
  \href{http://dx.doi.org/10.1103/PhysRevD.103.023527}{{\em Phys. Rev. D} {\bf
  103} (2021) no.~2, 023527}, \href{http://arxiv.org/abs/2005.05332}{{\tt
  arXiv:2005.05332 [hep-ph]}}.

\bibitem{Das:2022xsz}
A.~Das, Y.~F. Perez-Gonzalez, and M.~Sen, ``{Neutrino secret self-interactions:
  A booster shot for the cosmic neutrino background},''
  \href{http://dx.doi.org/10.1103/PhysRevD.106.095042}{{\em Phys. Rev. D} {\bf
  106} (2022) no.~9, 095042}, \href{http://arxiv.org/abs/2204.11885}{{\tt
  arXiv:2204.11885 [hep-ph]}}.

\bibitem{Akita:2022etk}
K.~Akita, S.~H. Im, and M.~Masud, ``{Probing non-standard neutrino interactions
  with a light boson from next galactic and diffuse supernova neutrinos},''
  \href{http://dx.doi.org/10.1007/JHEP12(2022)050}{{\em JHEP} {\bf 12} (2022)
  050}, \href{http://arxiv.org/abs/2206.06852}{{\tt arXiv:2206.06852
  [hep-ph]}}.

\bibitem{Balantekin:2023jlg}
A.~B. Balantekin, G.~M. Fuller, A.~Ray, and A.~M. Suliga, ``{Probing
  self-interacting sterile neutrino dark matter with the diffuse supernova
  neutrino background},''
  \href{http://dx.doi.org/10.1103/PhysRevD.108.123011}{{\em Phys. Rev. D} {\bf
  108} (2023) no.~12, 123011}, \href{http://arxiv.org/abs/2310.07145}{{\tt
  arXiv:2310.07145 [hep-ph]}}.

\bibitem{Doring:2023vmk}
C.~D\"oring and S.~Vogl, ``{Testing secret interaction with astrophysical
  neutrino point sources},''
  \href{http://dx.doi.org/10.1088/1475-7516/2024/07/015}{{\em JCAP} {\bf 07}
  (2024)  015}, \href{http://arxiv.org/abs/2304.08533}{{\tt arXiv:2304.08533
  [hep-ph]}}.

\bibitem{Huang:2017egl}
G.-y. Huang, T.~Ohlsson, and S.~Zhou, ``{Observational Constraints on Secret
  Neutrino Interactions from Big Bang Nucleosynthesis},''
  \href{http://dx.doi.org/10.1103/PhysRevD.97.075009}{{\em Phys. Rev. D} {\bf
  97} (2018) no.~7, 075009}, \href{http://arxiv.org/abs/1712.04792}{{\tt
  arXiv:1712.04792 [hep-ph]}}.

\bibitem{Luo:2020sho}
X.~Luo, W.~Rodejohann, and X.-J. Xu, ``{Dirac neutrinos and $N_{{\rm eff}}$},''
  \href{http://dx.doi.org/10.1088/1475-7516/2020/06/058}{{\em JCAP} {\bf 06}
  (2020)  058}, \href{http://arxiv.org/abs/2005.01629}{{\tt arXiv:2005.01629
  [hep-ph]}}.

\bibitem{Grohs:2020xxd}
E.~Grohs, G.~M. Fuller, and M.~Sen, ``{Consequences of neutrino self
  interactions for weak decoupling and big bang nucleosynthesis},''
  \href{http://dx.doi.org/10.1088/1475-7516/2020/07/001}{{\em JCAP} {\bf 07}
  (2020)  001}, \href{http://arxiv.org/abs/2002.08557}{{\tt arXiv:2002.08557
  [astro-ph.CO]}}.

\bibitem{Li:2023puz}
S.-P. Li and X.-J. Xu, ``{$N_{\rm eff}$ constraints on light mediators coupled
  to neutrinos: the dilution-resistant effect},''
  \href{http://dx.doi.org/10.1007/JHEP10(2023)012}{{\em JHEP} {\bf 10} (2023)
  012}, \href{http://arxiv.org/abs/2307.13967}{{\tt arXiv:2307.13967
  [hep-ph]}}.

\bibitem{Wu:2023twu}
Q.-f. Wu and X.-J. Xu, ``{Shedding light on neutrino self-interactions with
  solar antineutrino searches},''
  \href{http://dx.doi.org/10.1088/1475-7516/2024/02/037}{{\em JCAP} {\bf 02}
  (2024)  037}, \href{http://arxiv.org/abs/2308.15849}{{\tt arXiv:2308.15849
  [hep-ph]}}.

\bibitem{Akita:2023iwq}
K.~Akita, S.~H. Im, M.~Masud, and S.~Yun, ``{Limits on heavy neutral leptons,
  Z$^{'}$ bosons and majorons from high-energy supernova neutrinos},''
  \href{http://dx.doi.org/10.1007/JHEP07(2024)057}{{\em JHEP} {\bf 07} (2024)
  057}, \href{http://arxiv.org/abs/2312.13627}{{\tt arXiv:2312.13627
  [hep-ph]}}.

\bibitem{Wang:2023csv}
I.~R. Wang and X.-J. Xu, ``{Imprints of light dark matter on the evolution of
  cosmic neutrinos},''
  \href{http://dx.doi.org/10.1088/1475-7516/2024/05/050}{{\em JCAP} {\bf 05}
  (2024)  050}, \href{http://arxiv.org/abs/2312.17151}{{\tt arXiv:2312.17151
  [hep-ph]}}.

\bibitem{Kaplan:2024ydw}
D.~E. Kaplan, X.~Luo, and S.~Rajendran, ``{Probing Long-Range Forces Between
  Neutrinos with Cosmic Structures},''
  \href{http://arxiv.org/abs/2412.20766}{{\tt arXiv:2412.20766 [hep-ph]}}.

\bibitem{Horiuchi:2008jz}
S.~Horiuchi, J.~F. Beacom, and E.~Dwek, ``{The Diffuse Supernova Neutrino
  Background is detectable in Super-Kamiokande},''
  \href{http://dx.doi.org/10.1103/PhysRevD.79.083013}{{\em Phys. Rev. D} {\bf
  79} (2009)  083013}, \href{http://arxiv.org/abs/0812.3157}{{\tt
  arXiv:0812.3157 [astro-ph]}}.

\bibitem{Beacom:2010kk}
J.~F. Beacom, ``{The Diffuse Supernova Neutrino Background},''
  \href{http://dx.doi.org/10.1146/annurev.nucl.010909.083331}{{\em Ann. Rev.
  Nucl. Part. Sci.} {\bf 60} (2010)  439--462},
  \href{http://arxiv.org/abs/1004.3311}{{\tt arXiv:1004.3311 [astro-ph.HE]}}.

\bibitem{Fiorillo:2022cdq}
D.~F.~G. Fiorillo, G.~G. Raffelt, and E.~Vitagliano, ``{Strong Supernova 1987A
  Constraints on Bosons Decaying to Neutrinos},''
  \href{http://dx.doi.org/10.1103/PhysRevLett.131.021001}{{\em Phys. Rev.
  Lett.} {\bf 131} (2023) no.~2, 021001},
  \href{http://arxiv.org/abs/2209.11773}{{\tt arXiv:2209.11773 [hep-ph]}}.

\bibitem{KamLAND-Zen:2012uen}
{\bf KamLAND-Zen} Collaboration, A.~Gando {\em et al.}, ``{Limits on
  Majoron-emitting double-beta decays of Xe-136 in the KamLAND-Zen
  experiment},'' \href{http://dx.doi.org/10.1103/PhysRevC.86.021601}{{\em Phys.
  Rev. C} {\bf 86} (2012)  021601}, \href{http://arxiv.org/abs/1205.6372}{{\tt
  arXiv:1205.6372 [hep-ex]}}.

\bibitem{DESI:2024mwx}
{\bf DESI} Collaboration, A.~G. Adame {\em et al.}, ``{DESI 2024 VI:
  Cosmological Constraints from the Measurements of Baryon Acoustic
  Oscillations},'' \href{http://arxiv.org/abs/2404.03002}{{\tt arXiv:2404.03002
  [astro-ph.CO]}}.

\bibitem{Herold:2024nvk}
L.~Herold and M.~Kamionkowski, ``{Revisiting the impact of neutrino mass
  hierarchies on neutrino mass constraints in light of recent DESI data},''
  \href{http://arxiv.org/abs/2412.03546}{{\tt arXiv:2412.03546 [astro-ph.CO]}}.

\bibitem{Esteban:2024eli}
I.~Esteban, M.~C. Gonzalez-Garcia, M.~Maltoni, I.~Martinez-Soler, J.~a.~P.
  Pinheiro, and T.~Schwetz, ``{NuFit-6.0: updated global analysis of
  three-flavor neutrino oscillations},''
  \href{http://dx.doi.org/10.1007/JHEP12(2024)216}{{\em JHEP} {\bf 12} (2024)
  216}, \href{http://arxiv.org/abs/2410.05380}{{\tt arXiv:2410.05380
  [hep-ph]}}.

\bibitem{Wang:2024hen}
D.~Wang, O.~Mena, E.~Di~Valentino, and S.~Gariazzo, ``{Updating neutrino mass
  constraints with background measurements},''
  \href{http://dx.doi.org/10.1103/PhysRevD.110.103536}{{\em Phys. Rev. D} {\bf
  110} (2024) no.~10, 103536}, \href{http://arxiv.org/abs/2405.03368}{{\tt
  arXiv:2405.03368 [astro-ph.CO]}}.

\bibitem{Craig:2024tky}
N.~Craig, D.~Green, J.~Meyers, and S.~Rajendran, ``{No \ensuremath{\nu}s is
  Good News},'' \href{http://dx.doi.org/10.1007/JHEP09(2024)097}{{\em JHEP}
  {\bf 09} (2024)  097}, \href{http://arxiv.org/abs/2405.00836}{{\tt
  arXiv:2405.00836 [astro-ph.CO]}}.

\bibitem{Fardon:2003eh}
R.~Fardon, A.~E. Nelson, and N.~Weiner, ``{Dark energy from mass varying
  neutrinos},'' \href{http://dx.doi.org/10.1088/1475-7516/2004/10/005}{{\em
  JCAP} {\bf 10} (2004)  005},
  \href{http://arxiv.org/abs/astro-ph/0309800}{{\tt arXiv:astro-ph/0309800}}.

\bibitem{Kaplan:2004dq}
D.~B. Kaplan, A.~E. Nelson, and N.~Weiner, ``{Neutrino oscillations as a probe
  of dark energy},''
  \href{http://dx.doi.org/10.1103/PhysRevLett.93.091801}{{\em Phys. Rev. Lett.}
  {\bf 93} (2004)  091801}, \href{http://arxiv.org/abs/hep-ph/0401099}{{\tt
  arXiv:hep-ph/0401099}}.

\bibitem{Barger:2005mn}
V.~Barger, P.~Huber, and D.~Marfatia, ``{Solar mass-varying neutrino
  oscillations},'' \href{http://dx.doi.org/10.1103/PhysRevLett.95.211802}{{\em
  Phys. Rev. Lett.} {\bf 95} (2005)  211802},
  \href{http://arxiv.org/abs/hep-ph/0502196}{{\tt arXiv:hep-ph/0502196}}.

\bibitem{Cirelli:2005sg}
M.~Cirelli, M.~C. Gonzalez-Garcia, and C.~Pena-Garay, ``{Mass varying neutrinos
  in the sun},'' \href{http://dx.doi.org/10.1016/j.nuclphysb.2005.04.034}{{\em
  Nucl. Phys. B} {\bf 719} (2005)  219--233},
  \href{http://arxiv.org/abs/hep-ph/0503028}{{\tt arXiv:hep-ph/0503028}}.

\bibitem{Tait:2008zz}
T.~M.~P. Tait, ``{TASI Lectures on Resonances},''
\newblock 2009.
\newblock \url{{www.physics.uci.edu/~ttait/tait-TASI08.pdf}}.

\bibitem{Baldoncini:2014vda}
M.~Baldoncini, I.~Callegari, G.~Fiorentini, F.~Mantovani, B.~Ricci, V.~Strati,
  and G.~Xhixha, ``{Reference worldwide model for antineutrinos from
  reactors},'' \href{http://dx.doi.org/10.1103/PhysRevD.91.065002}{{\em Phys.
  Rev. D} {\bf 91} (2015) no.~6, 065002},
  \href{http://arxiv.org/abs/1411.6475}{{\tt arXiv:1411.6475
  [physics.ins-det]}}.

\bibitem{Vitagliano:2019yzm}
E.~Vitagliano, I.~Tamborra, and G.~Raffelt, ``{Grand Unified Neutrino Spectrum
  at Earth: Sources and Spectral Components},''
  \href{http://dx.doi.org/10.1103/RevModPhys.92.045006}{{\em Rev. Mod. Phys.}
  {\bf 92} (2020)  45006}, \href{http://arxiv.org/abs/1910.11878}{{\tt
  arXiv:1910.11878 [astro-ph.HE]}}.

\bibitem{Hyper-Kamiokande:2018ofw}
{\bf Hyper-Kamiokande} Collaboration, K.~Abe {\em et al.}, ``{Hyper-Kamiokande
  Design Report},'' \href{http://arxiv.org/abs/1805.04163}{{\tt
  arXiv:1805.04163 [physics.ins-det]}}.

\bibitem{Abe:2011ts}
K.~Abe {\em et al.}, ``{Letter of Intent: The Hyper-Kamiokande Experiment ---
  Detector Design and Physics Potential ---},''
  \href{http://arxiv.org/abs/1109.3262}{{\tt arXiv:1109.3262 [hep-ex]}}.

\bibitem{ParticleDataGroup:2024cfk}
{\bf Particle Data Group} Collaboration, S.~Navas {\em et al.}, ``{Review of
  particle physics},''
  \href{http://dx.doi.org/10.1103/PhysRevD.110.030001}{{\em Phys. Rev. D} {\bf
  110} (2024) no.~3, 030001}.

\bibitem{Sandner:2023ptm}
S.~Sandner, M.~Escudero, and S.~J. Witte, ``{Precision CMB constraints on
  eV-scale bosons coupled to neutrinos},''
  \href{http://dx.doi.org/10.1140/epjc/s10052-023-11864-6}{{\em Eur. Phys. J.
  C} {\bf 83} (2023) no.~8, 709}, \href{http://arxiv.org/abs/2305.01692}{{\tt
  arXiv:2305.01692 [hep-ph]}}.

\bibitem{Choi:1989hi}
K.~Choi and A.~Santamaria, ``{Majorons and Supernova Cooling},''
  \href{http://dx.doi.org/10.1103/PhysRevD.42.293}{{\em Phys. Rev. D} {\bf 42}
  (1990)  293--306}.

\bibitem{Kachelriess:2000qc}
M.~Kachelriess, R.~Tomas, and J.~W.~F. Valle, ``{Supernova bounds on Majoron
  emitting decays of light neutrinos},''
  \href{http://dx.doi.org/10.1103/PhysRevD.62.023004}{{\em Phys. Rev. D} {\bf
  62} (2000)  023004}, \href{http://arxiv.org/abs/hep-ph/0001039}{{\tt
  arXiv:hep-ph/0001039}}.

\bibitem{Farzan:2002wx}
Y.~Farzan, ``{Bounds on the coupling of the Majoron to light neutrinos from
  supernova cooling},''
  \href{http://dx.doi.org/10.1103/PhysRevD.67.073015}{{\em Phys. Rev. D} {\bf
  67} (2003)  073015}, \href{http://arxiv.org/abs/hep-ph/0211375}{{\tt
  arXiv:hep-ph/0211375}}.

\bibitem{Vogl:2024ack}
S.~Vogl and X.-J. Xu, ``{Heating the dark matter halo with dark radiation from
  supernovae},'' \href{http://arxiv.org/abs/2411.18052}{{\tt arXiv:2411.18052
  [hep-ph]}}.

\bibitem{Brune:2018sab}
T.~Brune and H.~P\"as, ``{Massive Majorons and constraints on the
  Majoron-neutrino coupling},''
  \href{http://dx.doi.org/10.1103/PhysRevD.99.096005}{{\em Phys. Rev. D} {\bf
  99} (2019) no.~9, 096005}, \href{http://arxiv.org/abs/1808.08158}{{\tt
  arXiv:1808.08158 [hep-ph]}}.

\bibitem{Blum:2018ljv}
K.~Blum, Y.~Nir, and M.~Shavit, ``{Neutrinoless double-beta decay with massive
  scalar emission},''
  \href{http://dx.doi.org/10.1016/j.physletb.2018.08.022}{{\em Phys. Lett. B}
  {\bf 785} (2018)  354--361}, \href{http://arxiv.org/abs/1802.08019}{{\tt
  arXiv:1802.08019 [hep-ph]}}.

\bibitem{GERDA:2022ffe}
{\bf GERDA} Collaboration, M.~Agostini {\em et al.}, ``{Search for exotic
  physics in double-\ensuremath{\beta} decays with GERDA Phase~II},''
  \href{http://dx.doi.org/10.1088/1475-7516/2022/12/012}{{\em JCAP} {\bf 12}
  (2022)  012}, \href{http://arxiv.org/abs/2209.01671}{{\tt arXiv:2209.01671
  [nucl-ex]}}.

\bibitem{Planck:2018vyg}
{\bf Planck} Collaboration, N.~Aghanim {\em et al.}, ``{Planck 2018 results.
  VI. Cosmological parameters},''
  \href{http://dx.doi.org/10.1051/0004-6361/201833910}{{\em Astron. Astrophys.}
  {\bf 641} (2020)  A6}, \href{http://arxiv.org/abs/1807.06209}{{\tt
  arXiv:1807.06209 [astro-ph.CO]}}. [Erratum: Astron.Astrophys. 652, C4
  (2021)].

\bibitem{Li:2022bpp}
S.-P. Li and X.-J. Xu, ``{Dark matter produced from right-handed neutrinos},''
  \href{http://dx.doi.org/10.1088/1475-7516/2023/06/047}{{\em JCAP} {\bf 06}
  (2023)  047}, \href{http://arxiv.org/abs/2212.09109}{{\tt arXiv:2212.09109
  [hep-ph]}}.

\bibitem{Schmitz:2023pfy}
K.~Schmitz and X.-J. Xu, ``{Wash-in leptogenesis after the evaporation of
  primordial black holes},''
  \href{http://dx.doi.org/10.1016/j.physletb.2024.138473}{{\em Phys. Lett. B}
  {\bf 849} (2024)  138473}, \href{http://arxiv.org/abs/2311.01089}{{\tt
  arXiv:2311.01089 [hep-ph]}}.

\bibitem{Suliga:2022ica}
A.~M. Suliga, {\em {Diffuse Supernova Neutrino Background}},
  \href{http://dx.doi.org/10.1007/978-981-15-8818-1_129-1}{pp.~1--18}.
\newblock 2022.
\newblock \href{http://arxiv.org/abs/2207.09632}{{\tt arXiv:2207.09632
  [astro-ph.HE]}}.

\bibitem{Horiuchi:2020jnc}
S.~Horiuchi, T.~Kinugawa, T.~Takiwaki, K.~Takahashi, and K.~Kotake, ``{Impact
  of binary interactions on the diffuse supernova neutrino background},''
  \href{http://dx.doi.org/10.1103/PhysRevD.103.043003}{{\em Phys. Rev. D} {\bf
  103} (2021) no.~4, 043003}, \href{http://arxiv.org/abs/2012.08524}{{\tt
  arXiv:2012.08524 [astro-ph.HE]}}.

\bibitem{Tabrizi:2020vmo}
Z.~Tabrizi and S.~Horiuchi, ``{Flavor Triangle of the Diffuse Supernova
  Neutrino Background},''
  \href{http://dx.doi.org/10.1088/1475-7516/2021/05/011}{{\em JCAP} {\bf 05}
  (2021)  011}, \href{http://arxiv.org/abs/2011.10933}{{\tt arXiv:2011.10933
  [hep-ph]}}.

\bibitem{Kresse:2020nto}
D.~Kresse, T.~Ertl, and H.-T. Janka, ``{Stellar Collapse Diversity and the
  Diffuse Supernova Neutrino Background},''
  \href{http://dx.doi.org/10.3847/1538-4357/abd54e}{{\em Astrophys. J.} {\bf
  909} (2021) no.~2, 169}, \href{http://arxiv.org/abs/2010.04728}{{\tt
  arXiv:2010.04728 [astro-ph.HE]}}.

\bibitem{Horiuchi:2017qja}
S.~Horiuchi, K.~Sumiyoshi, K.~Nakamura, T.~Fischer, A.~Summa, T.~Takiwaki,
  H.-T. Janka, and K.~Kotake, ``{Diffuse supernova neutrino background from
  extensive core-collapse simulations of $8$-$100 {\rm M}_\odot$
  progenitors},'' \href{http://dx.doi.org/10.1093/mnras/stx3271}{{\em Mon. Not.
  Roy. Astron. Soc.} {\bf 475} (2018) no.~1, 1363--1374},
  \href{http://arxiv.org/abs/1709.06567}{{\tt arXiv:1709.06567 [astro-ph.HE]}}.

\bibitem{Lunardini:2009ya}
C.~Lunardini, ``{Diffuse neutrino flux from failed supernovae},''
  \href{http://dx.doi.org/10.1103/PhysRevLett.102.231101}{{\em Phys. Rev.
  Lett.} {\bf 102} (2009)  231101}, \href{http://arxiv.org/abs/0901.0568}{{\tt
  arXiv:0901.0568 [astro-ph.SR]}}.

\bibitem{Yuksel:2008cu}
H.~Yuksel, M.~D. Kistler, J.~F. Beacom, and A.~M. Hopkins, ``{Revealing the
  High-Redshift Star Formation Rate with Gamma-Ray Bursts},''
  \href{http://dx.doi.org/10.1086/591449}{{\em Astrophys. J. Lett.} {\bf 683}
  (2008)  L5--L8}, \href{http://arxiv.org/abs/0804.4008}{{\tt arXiv:0804.4008
  [astro-ph]}}.

\bibitem{Super-Kamiokande:2021jaq}
{\bf Super-Kamiokande} Collaboration, K.~Abe {\em et al.}, ``{Diffuse supernova
  neutrino background search at Super-Kamiokande},''
  \href{http://dx.doi.org/10.1103/PhysRevD.104.122002}{{\em Phys. Rev. D} {\bf
  104} (2021) no.~12, 122002}, \href{http://arxiv.org/abs/2109.11174}{{\tt
  arXiv:2109.11174 [astro-ph.HE]}}.

\bibitem{Super-Kamiokande:2002weg}
{\bf Super-Kamiokande} Collaboration, Y.~Fukuda {\em et al.}, ``{The
  Super-Kamiokande detector},''
  \href{http://dx.doi.org/10.1016/S0168-9002(03)00425-X}{{\em Nucl. Instrum.
  Meth. A} {\bf 501} (2003)  418--462}.

\bibitem{HK_prj_reports}
\url{https://www-sk.icrr.u-tokyo.ac.jp/en/hk/report/}.

\bibitem{Beacom:2003nk}
J.~F. Beacom and M.~R. Vagins, ``{GADZOOKS! Antineutrino Spectroscopy with
  Large Water Cherenkov Detectors},''
  \href{http://dx.doi.org/10.1103/PhysRevLett.93.171101}{{\em Phys. Rev. Lett.}
  {\bf 93} (2004)  171101}, \href{http://arxiv.org/abs/hep-ph/0309300}{{\tt
  arXiv:hep-ph/0309300}}.

\bibitem{Super-Kamiokande:2021the}
{\bf Super-Kamiokande} Collaboration, K.~Abe {\em et al.}, ``{First gadolinium
  loading to Super-Kamiokande},''
  \href{http://dx.doi.org/10.1016/j.nima.2021.166248}{{\em Nucl. Instrum. Meth.
  A} {\bf 1027} (2022)  166248}, \href{http://arxiv.org/abs/2109.00360}{{\tt
  arXiv:2109.00360 [physics.ins-det]}}.

\bibitem{Super-Kamiokande:2023xup}
{\bf Super-Kamiokande} Collaboration, M.~Harada {\em et al.}, ``{Search for
  Astrophysical Electron Antineutrinos in Super-Kamiokande with 0.01\%
  Gadolinium-loaded Water},''
  \href{http://dx.doi.org/10.3847/2041-8213/acdc9e}{{\em Astrophys. J. Lett.}
  {\bf 951} (2023) no.~2, L27}, \href{http://arxiv.org/abs/2305.05135}{{\tt
  arXiv:2305.05135 [astro-ph.HE]}}.

\bibitem{Huang:2015hro}
K.~Huang, {\em {Measurement of the neutrino-oxygen neutral current
  quasi-elastic interaction cross-section by observing nuclear de-excitation
  $\gamma$-rays in the T2K experiment}}.
\newblock PhD thesis, Kyoto U., 2015.

\bibitem{Vogel:1999zy}
P.~Vogel and J.~F. Beacom, ``{Angular distribution of neutron inverse beta
  decay, anti-neutrino(e) + p ---\ensuremath{>} e+ + n},''
  \href{http://dx.doi.org/10.1103/PhysRevD.60.053003}{{\em Phys. Rev. D} {\bf
  60} (1999)  053003}, \href{http://arxiv.org/abs/hep-ph/9903554}{{\tt
  arXiv:hep-ph/9903554}}.

\bibitem{Strumia:2003zx}
A.~Strumia and F.~Vissani, ``{Precise quasielastic neutrino/nucleon
  cross-section},'' \href{http://dx.doi.org/10.1016/S0370-2693(03)00616-6}{{\em
  Phys. Lett. B} {\bf 564} (2003)  42--54},
  \href{http://arxiv.org/abs/astro-ph/0302055}{{\tt arXiv:astro-ph/0302055}}.

\bibitem{Li:2014sea}
S.~W. Li and J.~F. Beacom, ``{First calculation of cosmic-ray muon spallation
  backgrounds for MeV astrophysical neutrino signals in Super-Kamiokande},''
  \href{http://dx.doi.org/10.1103/PhysRevC.89.045801}{{\em Phys. Rev. C} {\bf
  89} (2014)  045801}, \href{http://arxiv.org/abs/1402.4687}{{\tt
  arXiv:1402.4687 [hep-ph]}}.

\bibitem{Li:2015kpa}
S.~W. Li and J.~F. Beacom, ``{Spallation Backgrounds in Super-Kamiokande Are
  Made in Muon-Induced Showers},''
  \href{http://dx.doi.org/10.1103/PhysRevD.91.105005}{{\em Phys. Rev. D} {\bf
  91} (2015) no.~10, 105005}, \href{http://arxiv.org/abs/1503.04823}{{\tt
  arXiv:1503.04823 [hep-ph]}}.

\bibitem{Li:2015lxa}
S.~W. Li and J.~F. Beacom, ``{Tagging Spallation Backgrounds with Showers in
  Water-Cherenkov Detectors},''
  \href{http://dx.doi.org/10.1103/PhysRevD.92.105033}{{\em Phys. Rev. D} {\bf
  92} (2015) no.~10, 105033}, \href{http://arxiv.org/abs/1508.05389}{{\tt
  arXiv:1508.05389 [physics.ins-det]}}.

\bibitem{Super-Kamiokande:2015xra}
{\bf Super-Kamiokande} Collaboration, Y.~Zhang {\em et al.}, ``{First
  measurement of radioactive isotope production through cosmic-ray muon
  spallation in Super-Kamiokande IV},''
  \href{http://dx.doi.org/10.1103/PhysRevD.93.012004}{{\em Phys. Rev. D} {\bf
  93} (2016) no.~1, 012004}, \href{http://arxiv.org/abs/1509.08168}{{\tt
  arXiv:1509.08168 [hep-ex]}}.

\bibitem{Super-Kamiokande:2021snn}
{\bf Super-Kamiokande} Collaboration, S.~Locke {\em et al.}, ``{New methods and
  simulations for cosmogenic induced spallation removal in
  Super-Kamiokande-IV},''
  \href{http://dx.doi.org/10.1103/PhysRevD.110.032003}{{\em Phys. Rev. D} {\bf
  110} (2024) no.~3, 032003}, \href{http://arxiv.org/abs/2112.00092}{{\tt
  arXiv:2112.00092 [hep-ex]}}.

\bibitem{Nairat:2024upg}
O.~Nairat, J.~F. Beacom, and S.~W. Li, ``{Neutron tagging can greatly reduce
  spallation backgrounds in Super-Kamiokande},''
  \href{http://dx.doi.org/10.1103/PhysRevD.111.023014}{{\em Phys. Rev. D} {\bf
  111} (2025) no.~2, 023014}, \href{http://arxiv.org/abs/2409.10611}{{\tt
  arXiv:2409.10611 [hep-ph]}}.

\bibitem{Moller:2018kpn}
K.~M\o{}ller, A.~M. Suliga, I.~Tamborra, and P.~B. Denton, ``{Measuring the
  supernova unknowns at the next-generation neutrino telescopes through the
  diffuse neutrino background},''
  \href{http://dx.doi.org/10.1088/1475-7516/2018/05/066}{{\em JCAP} {\bf 05}
  (2018)  066}, \href{http://arxiv.org/abs/1804.03157}{{\tt arXiv:1804.03157
  [astro-ph.HE]}}.

\bibitem{Zhou:2023mou}
B.~Zhou and J.~F. Beacom, ``{First detailed calculation of atmospheric neutrino
  foregrounds to the diffuse supernova neutrino background in
  Super-Kamiokande},''
  \href{http://dx.doi.org/10.1103/PhysRevD.109.103003}{{\em Phys. Rev. D} {\bf
  109} (2024) no.~10, 103003}, \href{http://arxiv.org/abs/2311.05675}{{\tt
  arXiv:2311.05675 [hep-ph]}}.

\bibitem{Super-Kamiokande:2013ufi}
{\bf Super-Kamiokande} Collaboration, H.~Zhang {\em et al.}, ``{Supernova Relic
  Neutrino Search with Neutron Tagging at Super-Kamiokande-IV},''
  \href{http://dx.doi.org/10.1016/j.astropartphys.2014.05.004}{{\em Astropart.
  Phys.} {\bf 60} (2015)  41--46}, \href{http://arxiv.org/abs/1311.3738}{{\tt
  arXiv:1311.3738 [hep-ex]}}.

\end{thebibliography}\endgroup

\clearpage
\newpage
\onecolumngrid


\setcounter{equation}{0}
\setcounter{figure}{0}
\setcounter{table}{0}
\setcounter{page}{1}
\setcounter{section}{0}

\makeatletter
\renewcommand{\thesection}{S\arabic{section}}
\renewcommand{\theequation}{S\arabic{equation}}
\renewcommand{\thefigure}{S\arabic{figure}}
\renewcommand{\thetable}{S\arabic{table}}

\renewcommand{\theHfigure}{S\arabic{figure}}
\renewcommand{\theHtable}{S\arabic{table}}
\renewcommand{\theHequation}{S\arabic{equation}}
\makeatother

\begin{center}
    \textbf{\large Widen the Resonance: Probe a New Regime of Neutrino Self-interactions
    in Astrophysical Neutrinos} \\ 
    \vspace{0.5cm}
    { \it \large Supplemental Material}\\ 
    \vspace{0.5cm}
    {Isaac R. Wang, Xun-Jie Xu, and Bei Zhou}
\end{center}

Here we provide material that is not needed in the main text but further supports our results and may help guide potential future developments.
In Sec.~\ref{sec:coll} and Sec.~\ref{sec:sol}, we derive the collision terms involved in our Boltzmann equation and introduce the discretization techniques for a numerical solution of the equation.
In Sec.~\ref{sec_SM_DSNBspec}, we present the calculation of the spectra of the diffuse supernova neutrino background (DSNB) and discuss the uncertainties by comparing our results with several recent calculations.
In Sec.~\ref{sec_SM_HyperK}, we introduce the Hyper-Kamiokande experiment and its detection of DSNB.
In Sec.~\ref{sec:robust}, we generalize our $\chi^2$ analysis and show that our result is robust against the uncertainties.

\section{Collision terms in the Boltzmann equation}
\label{sec:coll}

In this work, we are mainly concerned with two processes, $\nu_{\rm A}+\nu_{\rm C}\to\phi$
and $\phi\to2\nu_{\rm A}$. To simplify our notations, we denote the two
processes by $1+2\to3$ and $1+2\leftarrow3$, where the numbers allow
us to conveniently mark quantities of the corresponding particles.
For instance, $p_{1}$ and $E_{3}$ denote the momentum of the first
particle and the energy of the third particle in the two processes
respectively. 
When associating these numbers to actual particles, particles $1$ and $3$ always refer to $\nu_{\rm A}$ and $\phi$, respectively, while particle $2$ can be $\nu_{\rm C}$ or $\nu_{\rm A}$, depending on whether it is in $1+2\to3$ or $1+2\leftarrow3$, respectively.
For convenience, we also define the following notations:
\begin{align}
d\Pi\equiv\frac{d^{3}p}{2E(2\pi)^{3}}\thinspace,\ \ \widetilde{dp}\equiv\frac{d^{3}p}{(2\pi)^{3}}\thinspace.
\end{align}
As is well known, the collision terms of a Boltzmann equation typically
involve multi-dimensional phase space integrals. In this work, we
encounter the following integrals. 
\begin{align}
\Gamma_{\mathring{1}+2\to3} & \equiv\frac{1}{2E_{1}}\int d\Pi_{2}d\Pi_{3}f_{2}|{\cal M}|^{2}(2\pi\delta)^{4}\thinspace,\label{eq:-10-0}\\
\Gamma_{\mathring{1}+2\leftarrow3} & \equiv\frac{\xi}{2E_{1}}\int d\Pi_{2}d\Pi_{3}f_{3}|{\cal M}|^{2}(2\pi\delta)^{4}\thinspace,\label{eq:-11-0}\\
\Gamma_{1+2\to\mathring{3}} & \equiv\frac{1}{2E_{3}}\int d\Pi_{1}d\Pi_{2}f_{1}f_{2}|{\cal M}|^{2}(2\pi\delta)^{4}\thinspace,\label{eq:-12-0}\\
\Gamma_{1+2\leftarrow\mathring{3}} & \equiv\frac{1}{2E_{3}}\int d\Pi_{1}d\Pi_{2}|{\cal M}|^{2}(2\pi\delta)^{4}\thinspace,\label{eq:-13-0}
\end{align}
where $f_{i}$ denotes the phase space distribution of the $i$-th
particle, $|{\cal M}|^{2}$ is the squared amplitude of the process,
and $(2\pi\delta)^{4}\equiv(2\pi)^{4}\delta^{(4)}(p_{1}+p_{2}-p_{3})$.
In Eq.~\eqref{eq:-11-0}, we have introduced a parameter $\xi$ which
will be discussed later. In addition, we also add a small circle ($\mathring{\ }$) on the top of particle $1$ or $3$ to indicate that the process is focused on this particle. For instance, $\Gamma_{\mathring{1}+2\to3}$
means the absorption rate of particle $1$ via this process, while
$\Gamma_{1+2\to\mathring{3}}$ means the production rate of particle $3$. 

Note that in this work, the squared amplitude $|{\cal M}|^{2}$ is an energy-independent constant:
\[|{\cal M}|^{2}=g_{\nu}^{2}m_{\phi}^{2}\,.\] 
More generally, the squared amplitude
of a two-to-one or one-to-two process is usually a constant. Hence,
to make our calculation applicable to more general cases, below we
keep $|{\cal M}|^{2}$ without substituting its explicit form. 

Since $|\mathcal{M}|^{2}$ is a constant, the phase space integration
of Eq.~\eqref{eq:-13-0} can be straightforwardly performed, and we
arrive at 
\begin{align}
\Gamma_{1+2\leftarrow\mathring{3}}=\frac{|{\cal M}|^{2}}{16\pi E_{3}}\thinspace,\label{eq:12b3o}
\end{align}
which is exactly the boosted decay rate of particle $3$.

Eq.~\eqref{eq:-10-0} requires the specific form of $f_{2}$. Assuming
$f_{2}=\exp(-E_{2}/T)$, the phase space integration can also be performed
analytically---see Appendix~A.2. of Ref.~\cite{Li:2022bpp} for
more details.  The result reads
\begin{align}
\Gamma_{\mathring{1}+2\to3} & =\frac{T|{\cal M}|^{2}}{16\pi E_{1}^{2}}\exp\left[-\frac{m_{3}^{2}}{4Tp_{1}}\right].\label{eq:1o2-3}
\end{align}

Eq.~\eqref{eq:-12-0} involves two phase-space distributions, $f_{1}$
and $f_{2}$. If both are thermal distributions, the phase space integration
is also straightforward, and the result can be found in Ref.~\cite{Li:2022bpp}.
In our work, since we have a non-thermal neutrino flux scattering
off a thermal background, we need to perform this integration with
$f_{2}=\exp(-E_{2}/T)$ while $f_{1}$ maintains an arbitrary form.
In this circumstance, the integration cannot be fully carried out but can be reduced to a one-dimensional integral. More specifically,
we first integrate out $p_{2}$, leaving the energy-conservation part
of the $\delta$ function to be integrated out later. Then we integrate
out the azimuthal angle of $p_{1}$, leading to
\begin{align}
\Gamma_{1+2\to\mathring{3}}=\int\left(p_{1}^{2}dp_{1}f_{1}\right)f_{2}\frac{1}{8E_{1}E_{2}E_{3}}\frac{1}{2\pi}\delta(E_{1}+E_{2}-E_{3})|\mathcal{M}|^{2}p_{3}^{2}d\cos\theta_{13}\frac{dp_{3}}{dE_{3}}\thinspace.
\end{align}
The next step would be to integrate out $\cos\theta_{13}$. As long
as the energy of $E_{3}$ is kinematically allowed, there must be
a value of $\cos\theta_{13}$ that hits the peak of the $\delta$
function. Thus, we can safely integrate out the $\delta$ function with this polar angle, arriving at
\begin{align}
\Gamma_{1+2\to\mathring{3}} & =\int\left(p_{1}^{2}dp_{1}f_{1}\right)f_{2}\frac{1}{16\pi E_{1}E_{2}E_{3}}\frac{1}{\Delta}|\mathcal{M}|^{2}\thinspace,\\
\Delta & =\frac{p_{1}p_{3}}{|E_{3}-E_{1}|}\thinspace.
\end{align}
Using $|E_{3}-E_{1}|=E_{2}$ and $p_{1}=E_{1}$, we arrive at
\begin{align}
\Gamma_{1+2\to\mathring{3}}=\frac{g_{\nu}^{2}m_{\phi}^{2}}{16\pi E_{3}p_{3}}\int_{p_{1}^{{\rm min}}}^{p_{1}^{{\rm max}}}dp_{1}f_{1}\exp\left[-\frac{E_{3}-p_{1}}{T}\right].\label{eq:12-3o}
\end{align}
Here the minimal and maximal values of $p_{1}$ are determined from
energy-momentum conservation:
\begin{align}
p_{1}^{{\rm min}}=\frac{E_{3}-p_{3}}{2}\thinspace,\ \ p_{1}^{{\rm max}}=\frac{E_{3}+p_{3}}{2}\thinspace.
\end{align}

Using the same technique above, one can also reduce Eq.~\eqref{eq:-11-0}
to
\begin{align}
\Gamma_{\mathring{1}+2\leftarrow3}=\frac{g_{\nu}^{2}m_{\phi}^{2}}{16\pi E_{1}^{2}}\int_{E_{3}^{{\rm min}}}^{\infty}dE_{3}f_{3}\thinspace,\label{eq:1o2b3}
\end{align}
where
\begin{align}
E_{3}^{{\rm min}}=\frac{m_{\phi}^{2}}{4p_{1}}+p_{1}\thinspace.
\end{align}

The results of the above phase space integrals can be used to readily
obtain the mean free path $L_{{\rm mean}}$ in Eq.~\eqref{eq:mean}.
Since $\Gamma_{\mathring{1}+2\to3}$ is the absorption rate of particle
$1$, it can be physically interpreted as the probability of the particle
being absorbed by the medium per unit time, implying that before the absorption, it can propagate through a mean distance of $v\langle t\rangle$
where $v=1$ is the velocity for relativistic particle, and $\langle t\rangle=1/\Gamma_{\mathring{1}+2\to3}$.
Using the result in Eq.~\eqref{eq:1o2-3}, we obtain the UR result
in Eq.~\eqref{eq:mean}. For the NR case, we start from Eq.~\eqref{eq:-10-0}
where $f_{2}$ now takes a nonrelativistic distribution. Then the
$\int d\Pi_{2}f_{2}$ part can be approximately replaced with $n_{2}/(2m_{2})$,
independent of the specific form of $f_{2}$, and the remaining part
gives rise to a quantity proportional to the cross section $\sigma_{{\rm NR}}$.
Combined together, we arrive at the NR result in Eq.~\eqref{eq:mean}.

Let us finally comment on the $\xi$ factor in Eq.~\eqref{eq:-11-0}.
In this work, we take $\xi=2$ to account for the fact that each $\phi$
decay generates two $\nu_{\rm A}$. Nevertheless, we would like to mention
that in some new physics scenarios, it is possible to have $\xi=1$.
For instance, if $\phi$ is coupled to a standard model neutrino and
a dark fermion $\chi$ (see e.g.~\cite{Wang:2023csv}), then the
propagation of $\chi$ should adopt $\xi=1$ because each $\phi$
particle produce only one $\chi$ particle and each $\chi$ particle after scattering off a neutrino produces only one $\phi$ particle.
This would lead to the conservation of the total particle number,
which can be verified explicitly by the Boltzmann equation of their
comoving number densities:
\begin{equation}
\frac{d}{a^{3}dt}\left[\begin{array}{c}
n_{1}a^{3}\\
n_{3}a^{3}
\end{array}\right]=\left[\begin{array}{c}
-\int\widetilde{dp}_{1}f_{1}\Gamma_{\mathring{1}+2\to3}+\int\widetilde{dp}_{1}\Gamma_{\mathring{1}+2\leftarrow3}\\
\int\widetilde{dp}_{3}\Gamma_{1+2\to\mathring{3}}-\int\widetilde{dp}_{3}\Gamma_{1+2\leftarrow\mathring{3}}f_{3}
\end{array}\right].\label{eq:Boltzmann-n}
\end{equation}
Summing the two rows together and substituting Eqs.~\eqref{eq:-10-0}-\eqref{eq:-13-0}
into it, we get $d(n_{1}a^{3}+n_{3}a^{3})/dt=0$ if $\xi=1$. If $\xi=2$, the conservation of particle number does not hold,  but the total energy
density $\times a^{4}$ is approximately conserved if all species
are relativistic. 

\section{Solving the Boltzmann equation}
\label{sec:sol}

When solving the Boltzmann equation of a phase space function $f$,
it is useful to introduce the comoving momentum $\tilde{p}\equiv pa$
and the following variable transformation:
\begin{equation}
\left(\begin{array}{c}
t\\
p
\end{array}\right)\to\left(\begin{array}{c}
a\\
\tilde{p}
\end{array}\right),\ \ \ f\to\tilde{f}:\tilde{f}\left(a,\ \tilde{p}\right)=f\left(t,\ p\right)\thinspace,\label{eq:var-trans}
\end{equation}
This allows us to rewrite the standard form of the Boltzmann equation
\begin{equation}
\left[\frac{\partial}{\partial t}-Hp\frac{\partial}{p}\right]f\left(t,\ p\right)=C^{(f)}\thinspace\label{eq:Boltz-f-basic}
\end{equation}
into the following form (see Appendix B of \cite{Li:2022bpp} for
further details):
\begin{equation}
aH\frac{\partial}{\partial a}\tilde{f}\left(a,\ \tilde{p}\right)=\left.C^{(f)}\right|_{f\to\tilde{f},\ p\to\tilde{p}/a}\thinspace.\label{eq:Boltz-f-adv}
\end{equation}

In many problems, the right-hand side of Eq.~\eqref{eq:Boltz-f-adv}
can be written into the form of $Y-\tilde{f}X$ where $Y$ and $X$
are some functions of $a$ and $\tilde{p}$. Then Eq.~\eqref{eq:Boltz-f-adv}
can be solved analytically by computing some integrals of $X$ and
$Y$---see Sec. II of Ref.~\cite{Creque-Sarbinowski:2020qhz} and
Appendix B of Ref.~\cite{Schmitz:2023pfy} for further details.

\begin{figure}
\centering

\includegraphics[width=0.49\textwidth]{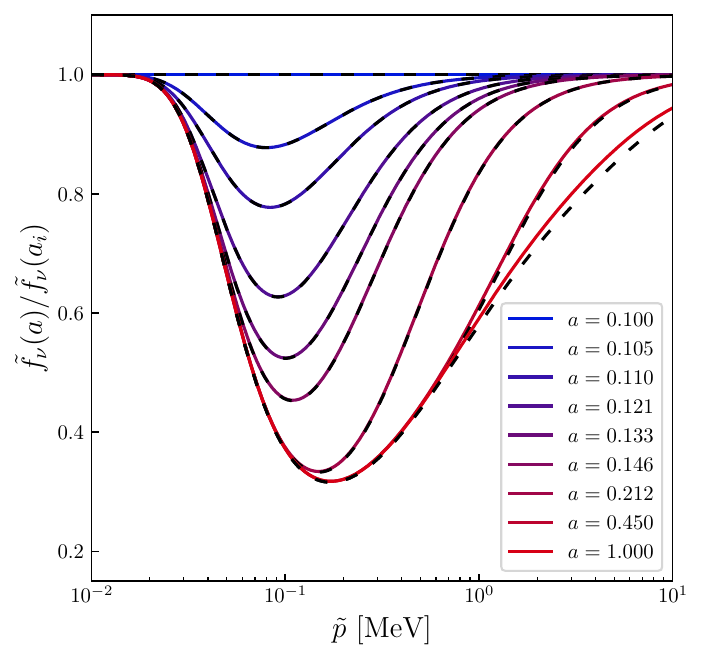}\includegraphics[width=0.49\textwidth]{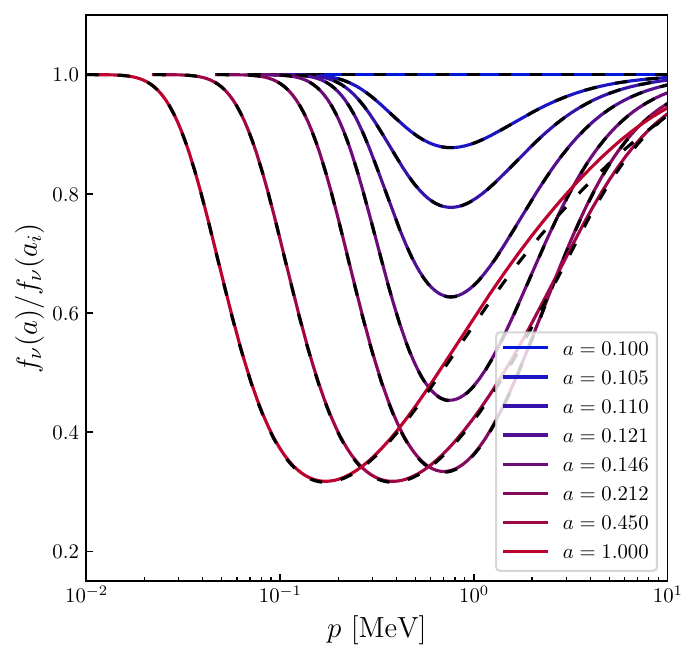}

\caption{Comparison of the analytical (dashed lines) and numerical (solid lines)
results of solving the Boltzmann equation of $f_{\nu}$, assuming
only the presence of the absorption term. In this sample, we set $m_{\phi}=10^{2}$
eV, $g=10^{-9}$, and $a_{i}=0.1$. \label{fig:solve-f}}
\end{figure}

In our work, this analytical approach allows us to compute $f_{\nu}$
if only $\nu_{\rm A}+\nu_{\rm C}\to\phi$ is taken into account while the backreaction is neglected (which is physically possible if $\phi$ has a dominant decay width to other final states). Following the analytical approach in Refs.~\cite{Creque-Sarbinowski:2020qhz,Schmitz:2023pfy},
we obtain the following analytical result in the absence of the backreaction:
\begin{equation}
\tilde{f}_{\nu}(a,\tilde{p})\approx\tilde{f}_{\nu}(a_{i},\tilde{p})\exp\left[\frac{a^{5/2}|{\cal M}|^{2}T_{0}}{32\pi\tilde{p}^{2}H_{0}x^{5/4}\Omega_{m}^{1/2}}\Gamma\left(\frac{5}{4},x\right)-C_{i}\right],\ \text{with}\ x\equiv\frac{a^{2}m_{\phi}^{2}}{4\tilde{p}T_{0}}\thinspace.\label{eq:f-ana}
\end{equation}
Here $a_{i}$ denotes the initial value of $a$, $T_{0}$ and $H_{0}$
denote the present value of $T$ and $H$, $\Gamma$ is an incomplete
gamma function, and $C_{i}$ is a constant that ensures the exponential
reduces to $1$ at $a=a_{i}$. When deriving this result, we have
assumed $H=H_{0}\sqrt{\Omega_{\Lambda}+\Omega_{m}/a^{3}}\approx H_{0}\sqrt{\Omega_{m}/a^{3}}$,
where $\Omega_{\Lambda}\approx0.692$ and $\Omega_{m}\approx0.308$ are the energy budgets of dark energy and matter in the present universe.
This approximation is valid for $10^{-3}\lesssim a\lesssim0.75$.
Figure~\ref{fig:solve-f} shows the comparison of the analytical result
in Eq.~\eqref{eq:f-ana} with the numerical solution (to be introduced
later). They are indeed in excellent agreement when the universe is matter dominated. 

In the presence of the backreaction, we need to solve Eq.~(\ref{eq:Boltzmann}),
which can only be solved numerically.  Our numerical approach is
built upon the discretized momentum space:
\begin{align}
\tilde{p} & \to\tilde{\boldsymbol{P}}\equiv\left(\tilde{p}_{1},\ \tilde{p}_{2},\ \tilde{p}_{3},\ \cdots\tilde{p}_{n}\right)^{T},\label{eq:-6}\\
\tilde{f} & \to F\equiv\left[\tilde{f}(\tilde{p}_{1}),\ \tilde{f}(\tilde{p}_{2}),\ \tilde{f}(\tilde{p}_{3}),\ \cdots,\tilde{f}(\tilde{p}_{n})\right]^{T}.
\end{align}
Here and below, we use capital or bold-font letters (e.g.~$\tilde{\boldsymbol{P}}$
and $F$) for some quantities to indicate that they are $n\times1$
column vectors. 

Next, we discretize and vectorize all momentum-dependent quantities
in Eq.~(\ref{eq:Boltzmann}), which can be rewritten into the following
matrix form: 
\begin{equation}
\frac{{\rm d}}{{\rm d}\ln a}\left[\begin{array}{c}
F_{\nu}\\
F_{\phi}
\end{array}\right]\approx\frac{1}{H}\left[\begin{array}{cc}
-B_{\mathring{1}+2\to3} & B_{\mathring{1}+2\leftarrow3}\\
B_{1+2\to\mathring{3}} & -B_{1+2\leftarrow\mathring{3}}
\end{array}\right]\left[\begin{array}{c}
F_{\nu}\\
F_{\phi}
\end{array}\right]+\frac{1}{H}\begin{bmatrix}S_{\nu}\\
0
\end{bmatrix}\thinspace,\label{eq:-18-1}
\end{equation}
where those $B$'s are $n\times n$ matrices to be elucidated later
and $S_{\nu}$ is the vectorized form of the source term:
\begin{align}
S_{\nu}\equiv\left[{\cal S}_{\nu}(\tilde{p}_{1}/a),{\cal S}_{\nu}(\tilde{p}_{2}/a),{\cal S}_{\nu}(\tilde{p}_{3}/a),\cdots,{\cal S}_{\nu}(\tilde{p}_{n}/a)\right]^{T}.
\end{align}

The $B$ matrices are obtained by discretizing Eqs.~\eqref{eq:12b3o}, \eqref{eq:1o2-3}, \eqref{eq:12-3o}, and \eqref{eq:1o2b3}:
\begin{align}
\left(B_{\mathring{1}+2\to3}\right)_{ii'} & \approx\left(\frac{T|{\cal M}|^{2}}{16\pi\boldsymbol{E}_{1}^{2}}\exp\left[-\frac{m_{3}^{2}}{4T\boldsymbol{E}_{1}}\right]\right)_{i}\delta_{ii'},\label{eq:-10-1}\\
\left(B_{\mathring{1}+2\leftarrow3}\right)_{ij} & \approx\left(\frac{\xi|{\cal M}|^{2}}{16\pi\boldsymbol{E}_{1}^{2}}\right)_{i}\Theta_{ij}^{(\text{cond.1})}\left(\boldsymbol{\Delta_{E_{3}}}\right)_{j}\thinspace,\label{eq:-11-1}\\
\left(B_{1+2\to\mathring{3}}\right)_{ji} & \approx\left(\frac{|{\cal M}|^{2}}{16\pi\boldsymbol{E}_{3}\boldsymbol{P}_{3}}e^{-\boldsymbol{E}_{3}/T}\right)_{j}\Theta_{ji}^{(\text{cond.2})}\left(\boldsymbol{\Delta_{P_{1}}}e^{\boldsymbol{P}_{1}/T}\right)_{i},\label{eq:-12-1}\\
\left(B_{1+2\leftarrow\mathring{3}}\right)_{jj'} & \approx\left(\frac{|{\cal M}|^{2}}{16\pi\boldsymbol{E}_{3}}\right)_{j}\delta_{jj'}\thinspace,\label{eq:-13-1}
\end{align}
where $\boldsymbol{\Delta_{X}}$ denotes the bin width of the its
subscript variable $\boldsymbol{X}$, corresponding to the discretized
form of $dX$. Note that here $\boldsymbol{E}$ and $\boldsymbol{P}$
are not comoving quantities so when the system evolves using the fixed
comoving grid of $\tilde{\boldsymbol{P}}$, they vary with $a$. 
The notation of $\Theta$ is defined as follows: 
\begin{equation}
\Theta_{ij}^{(\text{cond.})}=\begin{cases}
1 & \text{if cond.}=\text{true}\\
0 & \text{if cond.}=\text{false}
\end{cases}\thinspace,\label{eq:-17}
\end{equation} with two specific conditions: 
\begin{align}
\text{cond.1} & :\ \left(\frac{m_{3}^{2}}{4\boldsymbol{P}_{1}}+\boldsymbol{P}_{1}\right)_{i}<\left(\boldsymbol{E}_{3}\right)_{j}\thinspace,\label{eq:-25}\\
\text{cond.2} & :\ \left(\frac{\boldsymbol{E}_{3}-\boldsymbol{P}_{3}}{2}\right)_{j}<\left(\boldsymbol{P}_{1}\right)_{i}<\left(\frac{\boldsymbol{E}_{3}+\boldsymbol{P}_{3}}{2}\right)_{j}\thinspace.\label{eq:-26}
\end{align}
Assuming that $\phi$ is highly relativistic, we can further simplify
Eqs.~\eqref{eq:-11-1} and \eqref{eq:-12-1} into the following form:
\begin{align}
\left(B_{\mathring{1}+2\leftarrow3}\right)_{ij} & \approx\left(\frac{\xi|{\cal M}|^{2}}{16\pi\boldsymbol{E}_{1}^{2}}\right)_{i}\theta_{i<j}\left(\boldsymbol{\Delta_{E_{3}}}\right)_{j}\thinspace,\label{eq:-11-2}\\
\left(B_{1+2\to\mathring{3}}\right)_{ji} & \approx\left(\frac{|{\cal M}|^{2}}{16\pi\boldsymbol{E}_{3}\boldsymbol{P}_{3}}T\exp\left(-\frac{m_{3}^{2}}{4\boldsymbol{P}_{3}T}\right)\right)_{j}\delta_{ji}.\label{eq:-12-2}
\end{align}
where $\theta_{i<j}\equiv1$ if $i<j$ and $0$ if $i\geq j$. In
practice, we find that using Eqs.~\eqref{eq:-11-2} and \eqref{eq:-12-2}
Eqs.~\eqref{eq:-11-1} and \eqref{eq:-12-1} can greatly improve the
numeric stability and performance of an ODE solver. Besides, Eq.~\eqref{eq:-18-1}
allows one to conveniently implement the corresponding Jacobian to
be fed into an ODE solver. We find that this can improve the speed
of solving the equation typically by a factor of $10$. On a generic
laptop computer using the above method and the {\tt scipy.integrate.solve\_ivp}
solver in {\tt python} with the {\tt BDF} method, it typically takes ${\cal O}(0.1)$ seconds
to solve the discretized Boltzmann equation with $n=300$. 
A small {\tt atol} needs to be specified to achieve enough accuracy.

{
\section{DSNB spectra and uncertainties}
\label{sec_SM_DSNBspec}

The diffuse supernova neutrino background (DSNB) is the cumulative neutrino emission from core-collapse supernovae across the history of the Universe.
The DSNB neutrino flux can be calculated by~\cite{Horiuchi:2008jz,Beacom:2010kk,Suliga:2022ica}:
\begin{align}
\label{eq:dphidE}
    \frac{d \phi}{d E_\nu}(E_\nu) = \int \left((1+z) \frac{dN}{dE_\nu'}\right) R_{\rm SN}(z)  \frac{dt}{dz}dz\,,
\end{align}
where $E_\nu$ is the observed neutrino energy, $z$ the redshift, 
$dN/dE_\nu'$ the neutrino spectrum from a single supernova with $E_\nu' \equiv E_\nu (1+z)$ the neutrino energy at emission, and $R_{\rm SN}$ is the cosmic supernova explosion rate.
Moreover, unlike, e.g., Refs.~\cite{Horiuchi:2008jz,Beacom:2010kk}, we set $c=1$ in our equations to be consistent with the other equations in this paper.
Since Eq.~\eqref{eq:dphidE} is an integral over $z$ instead of time $t$,
it includes $dt/dz$, which can be determined from the Hubble rate, $H=a^{-1}da/dt$. 
Replacing $a$ with $1+z$, this implies $dz/dt=(1+z) H$. Due to the suppression of the star formation rate at high $z$, the DSNB calculation only requires performing the integration over the range $0 < z \lesssim 6$.
Within this range, the Hubble rate is approximately given by $H(z) = H_0 (\Omega_\Lambda + \Omega_m (1+z)^3)^{1/2}$.
We use $\Omega_\Lambda = 0.685$ and $\Omega_m = 0.315$~\cite{Planck:2018vyg, ParticleDataGroup:2024cfk}, and for the Hubble constant, we use $H_0 = 70~\rm km~s^{-1}~Mpc^{-1}$, consistent with the measurements in the local-Universe (e.g., using supernovae Ia)~\cite{ParticleDataGroup:2024cfk}, where DSNB are produced and propagate.

\begin{figure}
\centering
\includegraphics[width=0.9\textwidth]{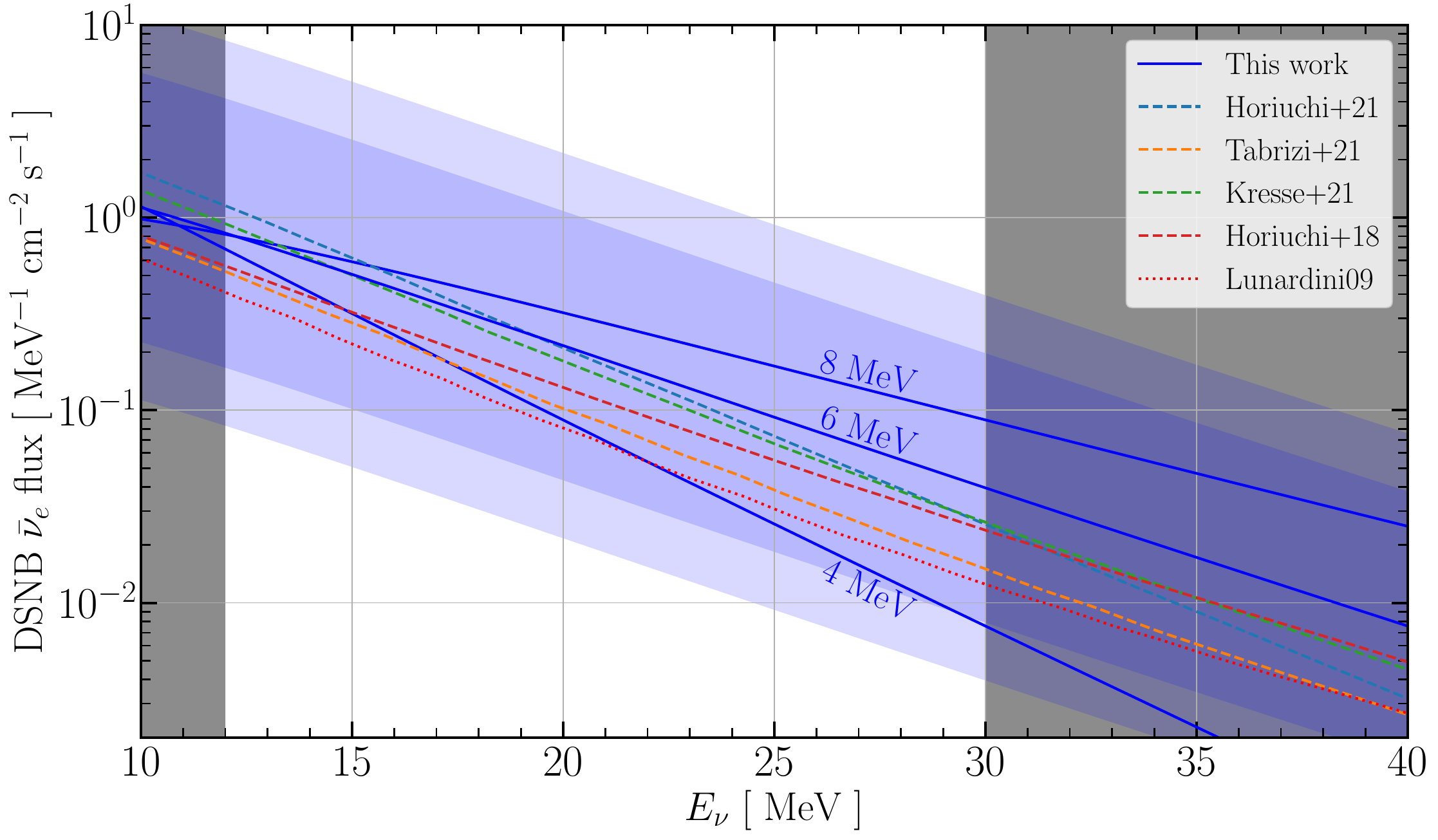}
\caption{
The DSNB spectra used in this work (blue lines, assuming $T_{\rm SN}=4,\ 6,\ 8$ MeV) compared with other calculations in the literature, including four published in recent years (Horiuchi+21~\cite{Horiuchi:2020jnc}, Tabrizi+21~\cite{Tabrizi:2020vmo}, Kresse+21~\cite{Kresse:2020nto}, Horiuchi+18~\cite{Horiuchi:2017qja}), and one known for its first inclusion of failed supernovae~(Lunardini09~\cite{Lunardini:2009ya}). 
The light blue shaded bands represent possible variations of the 6 MeV line
 assuming a conservative normalization uncertainty, $\sigma_\theta=0.5$ (see discussions in \ref{sec:robust}), with the thinner and thicker bands corresponding to $1 \sigma$ and $2 \sigma$ variations, respectively.
The gray regions correspond to $E_\nu < 12$~MeV and $E_\nu > 30$~MeV, which are beyond the data fitting window of our analysis due to high backgrounds (see Fig.~\ref{fig:events}).
}
\label{fig:fig_DSNB_spectra}
\end{figure}

To calculate Eq.~\eqref{eq:dphidE}, we need the supernova neutrino spectrum, $dN/dE_\nu'$, and the star formation rate, $R_{\rm SN}(z)$.
The former is usually assumed to follow Fermi-Dirac distribution~\cite{Horiuchi:2008jz,Beacom:2010kk}:
\begin{align}
    \frac{dN}{dE_\nu'}(E_\nu') = \frac{E_\nu^{\rm tot}}{6} \frac{120}{7 \pi^4} \frac{E_{\nu}'^2}{T_{\rm SN}^4} \left(e^{E_\nu'/T_{\rm SN} } + 1\right)^{-1}\,.
\end{align}
Here, $E_\nu^{\rm tot} = 3 \times 10^{53}~\rm erg$ is the total energy of neutrinos of all flavors emitted by a supernova, and 
$T_{\rm SN}$ is the DSNB temperature.
The star formation rate is given by~\cite{Yuksel:2008cu,Horiuchi:2008jz,Beacom:2010kk}
\begin{align}
    R_{\rm SN}(z) \simeq \dot{\rho}_*(z) \frac{0.01}{M_\odot}\,,
\end{align}
where
\begin{align}
    \dot{\rho}_*(z) = \dot{\rho}_0 \left((1+z)^{a \eta} + \left(\frac{1+z}{B}\right)^{b \eta} + \left(\frac{1+z}{C}\right)^{c \eta}\right)^{1/\eta}\,,
\end{align}
with the input parameters $a = 3.4$, $b=-0.3$, $c=-3.5$, $\eta = -10$, $B = 5161$, $C=9.06$, and $\dot{\rho}_0 = 0.0178 M_\odot~\rm yr^{-1}~ Mpc^{-3}$.

Fig.~\ref{fig:fig_DSNB_spectra} shows the DSNB flux of $\bar{\nu}_e$ calculated via the above procedure, assuming $T_{\rm SN}=4,\ 6,\ 8$ MeV, respectively.
Also shown are a few results from calculations within the recent ten years~\cite{Horiuchi:2017qja, Kresse:2020nto, Horiuchi:2020jnc, Tabrizi:2020vmo} as well as an early one from Ref.~\cite{Lunardini:2009ya}, known for its first inclusion of failed supernovae.
We refer readers to Fig.~1 of Ref.~\cite{Super-Kamiokande:2021jaq} for a more comprehensive compilation of DSNB calculations that include earlier works.
From Fig.~\ref{fig:fig_DSNB_spectra}, one can see that calculations in the literature in the recent ten years vary within about a factor of 3. This is consistent with the normalization uncertainty $\sigma_y=0.5$ used in the main text. In Sec.~\ref{sec:robust} below, we discuss a more conservative assumption on the normalization, parametrized by $\theta\equiv \lg(1+y)$ with the normalization ($1+y$) varying by one order of magnitude.
This corresponds to the blue shaded bands in Fig.~\ref{fig:fig_DSNB_spectra}. As is shown in the figure, the blue bands are sufficiently wide to cover the DSNB curves from different calculations. 
It also shows that our spectrum with $T=6$ MeV agrees with those calculated in the literature, while the interval of $[4,\ 8]$ MeV suffices to cover the spectral shape variation.

Finally, we discuss the impact of failed supernovae on the DSNB flux calculation.
Failed supernovae, also known as dark collapses, are core-collapse events in which the stellar core forms a black hole without producing a visible supernova, in contrast to those with successful explosions that typically result in neutron stars and rich optical signals.
Ref.~\cite{Lunardini:2009ya} was the first to include this correction, shown as the Lunardini09 line in Fig.~\ref{fig:fig_DSNB_spectra}; it is also included in the Horiuchi+18, Kresse+21, and Tabrizi+21 lines.
Comparing these lines and the others that do not include the failed supernova correction, including this component would slightly reduce the DSNB flux at lower energies while enhancing it above $\simeq30$~MeV. 
However, this component has large uncertainties and, as shown in Fig.~\ref{fig:fig_DSNB_spectra}, its effect on the DSNB spectral shape is much smaller than the variation in the DSNB temperature, i.e., from 4 to 8~MeV, in our calculations. 
Therefore, including failed supernovae does not significantly alter our results. 
}

\section{Hyper-Kamiokande and Its Detection of DSNB}
\label{sec_SM_HyperK}

For the detection of DSNB, we focus on Hyper-Kamiokande (HK)~\cite{Abe:2011ts, Hyper-Kamiokande:2018ofw}, a next-generation water Cherenkov detector in Japan that aims to achieve multiple physics goals including observations of neutrinos from astrophysical sources. As the successor to Super-Kamiokande~\cite{Super-Kamiokande:2002weg}, it will scale up the fiducial volume to 187 kton per tank, along with many other improvements such as electronics. 
Currently, one tank has been approved and is under construction, with data taking expected to begin in 2027~\cite{HK_prj_reports}, and a second tank is also under consideration.
Such a large fiducial volume makes HK one of the most promising detectors for DSNB detection.

The primary detection channel for DSNB in HK is the inverse beta decay (IBD), i.e., 
\begin{equation}
\bar{\nu}_e + p \rightarrow e^+ + n \,.
\end{equation}
This interaction is ideal for DSNB detection due to its relatively large cross section and the distinct event signature: a prompt positron signal followed by a delayed neutron capture. The latter can be significantly improved by gadolinium (Gd) doping. 
With a concentration of 0.1\% Gd by mass~\cite{Beacom:2003nk, Super-Kamiokande:2021the, Super-Kamiokande:2023xup}, HK will gain a much stronger neutron tagging capability via the $\simeq8$~MeV gamma-ray cascade from neutron capture on Gd.
The delayed-coincidence signature between the prompt positron and the neutron capture allows for powerful background suppression.

The expected DSNB event rate in HK can be calculated from a well established procedure\,---\,see e.g.~\cite{Balantekin:2023jlg} for useful details. More specifically, one can formulate the event rate as follows:
\begin{equation}
\frac{\dd N}{\dd E_{e}} = \epsilon N_t \Delta t \int \dd E_{\nu} \, \sigma_{\mathrm{IBD}}(E_{e},  E_{\nu}) \, \phi_{\bar{\nu}_e}(E_{\nu}) \,,
\label{eq_SM_DSNB_rate}
\end{equation}
where $\epsilon = 67\%$ is the detection efficiency~\cite{Huang:2015hro,Hyper-Kamiokande:2018ofw},
$N_t$ the number of target particles (for IBD, it is the number of hydrogen nuclei) in the fiducial volume,  
$\Delta t$ the exposure time, 
$\sigma_{\mathrm{IBD}}$ the IBD cross section, 
and $\phi_{\bar{\nu}_e}$ is the DSNB $\bar{\nu}_e$ flux. 
We consider a total exposure of $3740 ~{\rm kton}\cdot{\rm yr}$, which can be achieved by a single tank with 20 years of data taking, or by double tanks with 10 years. 
The IBD cross section is, in principle, a function of the positron energy $E_e$ and the neutrino energy $E_{\nu}$. 
However, as an excellent approximation due to the heavy masses of nucleons compared to $E_e$ and $E_{\nu}$, we use the one-to-one correspondence $E_{e} \simeq \left[\left(E_{\nu}-1.3\, \mathrm{MeV}\right)\right]\left(1-E_{\nu} / m_p\right)$, where the 1.3~MeV is the mass difference between a neutron and a proton and $m_p$ is the proton mass. 
The specific values of $\sigma_{\mathrm{IBD}}$ can be attained from Refs.~\cite{Vogel:1999zy, Strumia:2003zx}.

\begin{figure}
\centering
\includegraphics[width=0.8\textwidth]{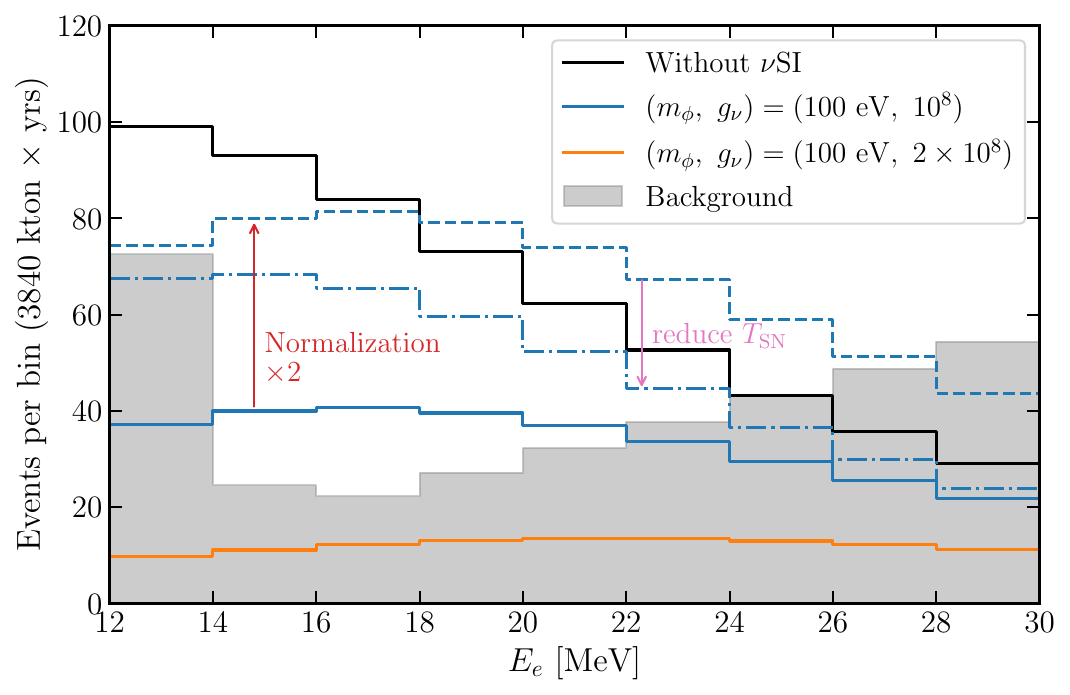}
\caption{The expected numbers of DSNB events at HK. The black line represents results without new physics involved. The blue and orange lines demonstrate the results in new physics scenarios with $m_{\phi}=100$ eV, $g_\nu=10^{-8}$ and $2\times 10^{-8}$, respectively. The dashed blue line is obtained by multiplying the solid blue line by a factor of two, and the dot-dashed line is obtained by reducing the DSNB temperature $T_{\rm SN}$. 
All shown examples take $T_{\rm SN}=6\, {\rm MeV}$, except for the dot-dashed line, which takes $T_{\rm SN}=5 {\rm MeV}$.
\label{fig:events}
}
\end{figure}

Fig.~\ref{fig:events} shows the event numbers of DSNB in HK calculated using Eq.~\eqref{eq_SM_DSNB_rate}, presented in 2-MeV bins since the energy resolution of HK in this energy range is around 2 MeV~\cite{Hyper-Kamiokande:2018ofw}. The energy window used in our analysis is set as [12, 30] MeV.  Within this window, we have a prominent signal-to-background ratio. 
The black line represents results without new physics involved. The blue and orange lines represent the results in new physics scenarios with $m_{\phi}=100$ eV, $g_\nu=10^{-8}$ and $2\times 10^{-8}$, respectively. The dashed and dot-dashed lines will be discussed in Sec.~\ref{sec:robust}. 
The gray histogram indicates the background level for DSNB detection in HK, taken from Ref.~\cite{Hyper-Kamiokande:2018ofw}.

Fig.~\ref{fig:events} also shows the backgrounds for DSNB detection in HK~\cite{Hyper-Kamiokande:2018ofw}.
They include:
\begin{enumerate}
\vspace{-0.2cm}
\item \textbf{invisible muons}, from atmospheric $\nu_{\mu}$ charged-current (CC) interactions and all-flavor atmospheric neutrino NC interactions with an invisible $\pi^+$ produced, 
\vspace{-0.2cm}
\item \textbf{the $\nu_e$CC component}, from atmospheric electron neutrino CC interactions, 
\vspace{-0.2cm}
\item \textbf{spallation background}, dominated by the $^9$Li component, from cosmic-ray muon interactions~\cite{Li:2014sea, Li:2015kpa, Li:2015lxa, Super-Kamiokande:2015xra, Super-Kamiokande:2021snn, Nairat:2024upg}, 
\vspace{-0.2cm}
\item \textbf{$\bar{\nu}_e$ from reactors}.
\end{enumerate}
\vspace{-0.2cm}
The DSNB detection backgrounds we use are very similar to those in, e.g., Refs.~\cite{Moller:2018kpn, Balantekin:2023jlg}.

We refer to Ref.~\cite{Zhou:2023mou} for further explanations of the physics behind these backgrounds, particularly the invisible muon and the $\nu_e$CC components, see Ref.~\cite{Zhou:2023mou}.
The backgrounds from Ref.~\cite{Hyper-Kamiokande:2018ofw} for HK are essentially unaffected by uncertainties in the atmospheric neutrino flux and neutrino cross sections, as they are constructed based on actual measurements in Super-Kamiokande.

Comparing the signal and background in Fig.~\ref{fig:events}, the optimal energy window for DSNB detection is expected to be between approximately 12 and 30~MeV. We use this range for our analysis.

Last but not least, we believe that the backgrounds could be further reduced in the future analysis of HK data, due to the following considerations. 
1) With time, more methods to further reduce the backgrounds will be proposed and applied, resulting in lower backgrounds than those used in this work from Ref.~\cite{Hyper-Kamiokande:2018ofw}. 
2) We only consider the case of neutron captures on Gd (with 90\% probability) with identifiable delayed-coincidence gamma rays, but the rest of the captures on protons, which emit a 2.2~MeV gamma ray, are also detectable, with a lower detection efficiency. This has already been used in Super-Kamiokande analyses~\cite{Super-Kamiokande:2013ufi, Super-Kamiokande:2021jaq}.
3) The DSNB IBD events without neutron-capture detection can also be used, despite higher backgrounds, which have already been used in the most recent Super-Kamiokande analyses~\cite{Super-Kamiokande:2021jaq}.

{

\section{On the robustness of the statistical significance\label{sec:robust}}

As neutrino self-interactions may lead to a significant reduction in event rates, one may wonder whether normalization and temperature uncertainties could, to some extent, compensate for the reduction, and thereby diminish the statistical significance of the new physics effect.  To illustrate the impact of normalization and temperature uncertainties, we add a dashed line and a dot-dashed line in Fig.~\ref{fig:events}, obtained by changing the normalization factor and the DSNB temperature $T_{\rm SN}$ in the way indicated in the plot, assuming  $(m_{\phi},\ g_{\nu})=(100 ~{\rm eV}, 10^{-8})$. 
The dashed blue line shows that the reduction caused by neutrino self-interactions could be compensated by increasing the normalization, but the spectral distortion would still make it distinguishable from the case without neutrino self-interactions. Reducing the temperature $T_{\rm SN}$ could alleviate the tension of spectral distortion, at the cost of reducing the total event number. 
Overall, we find that varying $T_{\rm SN}$ and the normalization always yields $\chi^2\gtrsim 11.6$ (i.e., excluded at $>3\sigma$ C.L.) for this sample. Hence $(m_{\phi},\ g_{\nu})=(100 ~{\rm eV}, 10^{-8})$ is indeed within the sensitivity reach of HK. The orange curve shows that for $(m_{\phi},\ g_{\nu})=(100 ~{\rm eV}, 2\times 10^{-8})$, the DSNB would be reduced to an undetectable level for HK. Therefore, if the DSNB is successfully detected by HK, this new physics scenario can be easily excluded.

Our main result is robust even under the most conservative assumptions on the DSNB flux uncertainties. To demonstrate this, we allow the normalization factor to vary over a much wider range (e.g., one or two orders of magnitude) and investigate its impact on the result. 
The $1+y$ parametrization used in Eq.~\eqref{eq:chi} is not applicable to very large flux uncertainties. To account for potentially large variations within one or two orders of magnitude, we make the replacement $1+y\to10^{\theta}$, with $\theta$ varying in a relatively large range. Correspondingly, the $\chi^{2}$ function in this case is formulated as follows: 
\begin{equation}
\chi^{2}=\left(\frac{\theta}{\sigma_{\theta}}\right)^{2}+\sum_{i}\left(\frac{10^{\theta}N_{i}-N_{{\rm st},i}}{\sigma_{i}}\right)^{2}.\label{eq:chi-theta}
\end{equation}
The variation of $1+y\in[1-0.5,\ 1+0.5]$ approximately corresponds
to $\theta\in[-0.3,\ 0.2]$. So we expect that $\sigma_{\theta}=0.2$
should approximately reproduce the result of $\sigma_{y}=0.5$ used
in the main text.
On the other hand, increasing $\sigma_{\theta}$ to $0.5$ corresponds
to the normalization varying by one order of magnitude at $1\sigma$ (68\%)
C.L.
We stress here that this is a very conservative assumption on the flux uncertainty, as it only requires that 68\% of the calculated fluxes vary within one order of magnitude, and 95\% of them within
two orders of magnitude. According to discussions in Sec.~\ref{sec_SM_DSNBspec} and Fig.~\ref{fig:fig_DSNB_spectra},
DSNB predictions in recent years have converged into a relatively
narrow range of a factor of three, much smaller than the $1\sigma$ band of $\sigma_{\theta}=0.5$.
Hence, $\sigma_{\theta}=0.5$ should only be regarded as a demonstration
of the robustness of the statistical significance.  

Fig.~\ref{fig:chi2-ctf} shows the values of $\chi^2$ with $\sigma_y=0.5$, $\sigma_\theta = 0.2$, and $\sigma_\theta = 0.5$.
The first two panels present very similar results, confirming the expectation that $\sigma_\theta = 0.2$ should be roughly equivalent to $\sigma_y = 0.5$.
Meanwhile, the right panel demonstrates that the result
remains largely unaffected even for the flux uncertainty as large as $\sigma_\theta = 0.5$.
The most significant change appears at $m_\phi \geq 200~\rm eV$.
In this regime, the absorption 
causes less tilting of the spectrum within the observation window.
Consequently, its effect can be more easily compensated by flux normalization, leading to a weakened sensitivity.

\begin{figure}
\centering

\includegraphics[width=0.98\textwidth]{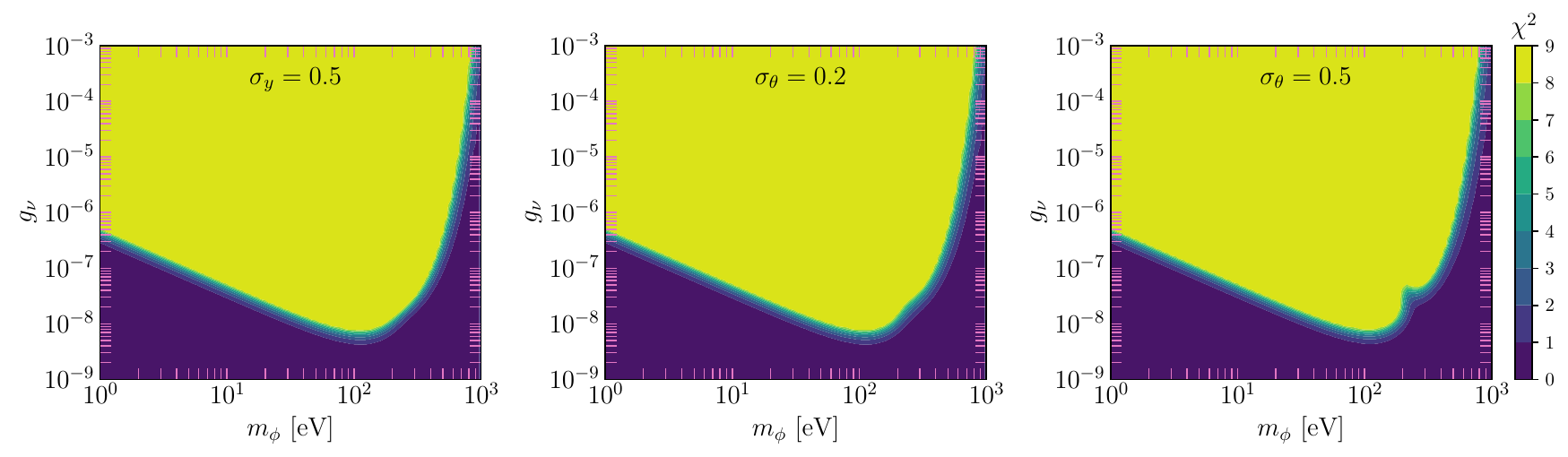}

\caption{The values of $\chi^{2}$ calculated using the $1+y$ parametrization
with $\sigma_{y}=0.5$ (left panel) and the $10^{\theta}$ parametrization
with $\sigma_{\theta}=0.2$ (middle panel) and $0.5$ (right panel).
\label{fig:chi2-ctf}
}

\end{figure}

The robustness of our result can be understood from the fact that our result is exponentially sensitive to the coupling $g_{\nu}$.
Taking $m_{\phi}=100\ {\rm eV}$ and $g_{\nu}=10^{-8}$ for example, which according to Fig.~\ref{fig:DSNB} would reduce the DSNB flux by $\sim 50\%$ (see also Fig.~\ref{fig:events} for the event numbers). If we increase $g_{\nu}$ to $2\times 10^{-8}$, the mean free path $L_{\rm mean}$ in Eq.~\eqref{eq:L-value} would be reduced by a factor of four, implying that the DSNB flux would be reduced by a factor of $e^{-4}\Phi_{\nu}\approx0.018$. 
Therefore, if we consider the abortion effect only,  $g_{\nu}=2\times 10^{-8}$ in this case would lead to \textit{a drastic reduction by two orders of magnitude}. 
After including the compensation from the regeneration effect, the flux is still reduced by one order of magnitude, implying that the observation of such a new physics effect should be statistically robust.

}

\end{document}